\documentclass[aps,prl,twocolumn,superscriptaddress,groupedaddress]{revtex4}
\usepackage[compat=1.0.0]{tikz-feynman}
\usepackage{amssymb}
\usepackage{multirow}
\usepackage{bm}
\usepackage{amsmath}
\usepackage{graphicx}
\usepackage{epstopdf}
\usepackage{subfigure}
\usepackage{natbib}
\usepackage{epsfig}
\usepackage{amsfonts}
\usepackage{mathrsfs}
\usepackage[toc,page,title,titletoc,header]{appendix}
\usepackage[colorlinks,linkcolor=blue,citecolor=blue,anchorcolor=blue]{hyperref}
\usepackage{dsfont,amsthm,amsbsy}

\usepackage{fancyhdr}

\usetikzlibrary{quotes,angles}
\newcommand{\obtuseangle}{\kern.08em
\begin{tikzpicture}
    \draw coordinate (a) at (0.14,0);
    \draw coordinate (b) at (0,0);
    \draw coordinate (c) at (-.12,0.18);
    \draw (a) -- (b) -- (c) pic [draw=black]{} ;
\end{tikzpicture}%
\kern.08em%
}

\begin{document}
\title{Strong Coupling Limit of the Holstein-Hubbard Model}
\author{Zhaoyu Han}
 \affiliation{Department of Physics, Stanford University, Stanford, California 94305, USA}

\author{Steven A. Kivelson}
\email{kivelson@stanford.edu}
  \affiliation{Department of Physics, Stanford University, Stanford, California 94305, USA}

\author{Hong Yao}
\email{yaohong@tsinghua.edu.cn}
 \affiliation{Institute of Advanced Study, Tsinghua University, Beijing 100084, China}
 \affiliation{Department of Physics, Stanford University, Stanford, California 94305, USA}

\date{\today}

\begin{abstract}
We analyze the quantum phase diagram of the Holstein-Hubbard model using an asymptotically exact strong-coupling expansion. We find all sorts of interesting phases including a pair-density wave (PDW), a charge 4e (and even a charge 6e) superconductor, regimes of phase separation, and a variety of distinct charge-density-wave (CDW), spin-density-wave (SDW) and superconducting regimes.  We chart the crossovers that occur as a function of the degree of retardation, i.e. the ratio of characteristic phonon frequency to the strength of interactions. 
\end{abstract}

\maketitle

Typically, in strongly correlated materials, both direct electron-electron interactions and electron-phonon interactions are strong.  None-the-less, most theoretical studies focus exclusively on one or the other. The most widely studied model  of the interplay between electron-electron (e-e) and electron-phonon (e-ph) interactions is the Holstein-Hubbard model~\cite{PhysRevLett.69.1600,PhysRevB.48.3881,PhysRevB.50.6939, fehske1994polaron,PhysRevB.52.4806, PhysRevLett.75.2570,PhysRevB.53.9666,Stein_1997,PROVILLE1998307,PhysRevB.64.094507,PhysRevB.67.081102,PhysRevLett.92.106401,PhysRevB.69.245111,  PhysRevB.70.155103, Koller_2004,Carlson2004, PhysRevLett.97.056402,PhysRevLett.99.146404, PhysRevB.75.245103, Gunnarsson_2008,PhysRevB.78.155123, PhysRevB.87.075149, PhysRevB.87.235133,PhysRevB.88.165108,  PhysRevLett.119.197001,PhysRevB.96.205145,PhysRevB.98.085405,hohenadler2018density,PhysRevB.98.201108,wang2019zero,kazuhiro2020phase}. The majority of existing studies are numerical explorations, despite the fact that the problem is complicated 
by the existence of multiple energy scales and a large parameter space.  Monte-Carlo studies on this problem are also generically rendered difficult by the fermion minus sign problem \cite{Yao-review}. In this letter we systematically explore the
``strong-coupling'' regimes in which the interactions are larger than the band-width, and a variety of results are derived from a theoretically well-controlled perturbative expansion.  Qualitative results are summarized in the schematic phase diagram in Fig.~\ref{pd}.

The Holstein-Hubbard model is defined as 
\begin{align}\label{Hamiltonian}
\hat{H} = & - t \sum_{\langle i,j\rangle,\sigma} (\hat{c
}_{i,\sigma}^{\dagger} \hat{c}_{j,\sigma}+ \text{h.c.}) -\mu\sum_j\hat n_j \nonumber\\
&+ \frac{U_{\text{e-e}}}{2} \sum_i \hat{n}_i^2 + \alpha \sum_i \hat{n}_i \hat x_i + \sum_i \left[ \frac{\hat{p}_i^2}{2m} + \frac{k\hat{x}_i^2}{2}\right]
\end{align}
where $\langle i,j\rangle$ signifies pairs of nearest-neighbor sites, $\hat{c}_{j\sigma}$ annihilates an electron with spin polarization $\sigma$ on site $j$, $\hat n_j= \sum_\sigma \hat{c}_{j\sigma}^\dagger \hat{c}_{j\sigma}$ is the number operator on site $j$,  $x_j$ is an optical phonon coordinate at site $j$ and $p_j$ is the conjugate momentum. The dominant effects of strong electron-phonon coupling can be accounted for by a unitary transformation $\hat{U} \equiv \prod_i \exp{\left[\mathrm{i}(\alpha/k)\hat{p}_i\hat{n}_i\right]}$~\cite{Hohenadler2007}. The transformed Hamiltonian is
\begin{align} \label{transformed}
\hat{U}^{\dagger}\hat{H} \hat{U} =& - t \sum_{\langle i,j\rangle ,\sigma} (\hat{S}_{ij}\hat{c}_{i,\sigma}^{\dagger} \hat{c}_{j,\sigma} + \text{h.c.}) -\mu\sum_j\hat n_j\nonumber\\
&+  \frac{U_{\text{eff}}}{2} \sum_i \hat{n}_i^2 + \sum_i \left[ \frac{\hat{p}_i^2}{2m} + \frac{k\hat{x}_i^2}{2}\right], 
\end{align}
where $\hat{S}_{ij} \equiv  \exp{\left[\mathrm{i}(\alpha/k)(\hat{p}_j - \hat{p}_i)\right]}$ is a product of two phonon displacement operators on site $i$ and $j$, $U_{\text{eff}}\equiv U_{\text{e-e}} - U_{\text{e-ph}}$, and $U_{\text{e-ph}} \equiv \alpha^2/k$. This transformation is exact and can be alternatively derived by a path integral representation tracking the phonon degrees of freedom in  momentum space~\footnote{Using this approach, we also prove in Supplemental Material that for $U_{\text{eff}}\geq 0$, the half-filled Holstein-Hubbard model is fermion sign problem free on bipartite lattice. This is a recently known conclusion~\cite{PhysRevB.98.201108}. }. 

In the strong coupling expansion, we treat the hopping term in the transformed Hamiltonian as a perturbation, and the sign of $U_{\text{eff}}$ determines the relevant low-energy degrees of freedom. The resulting theories are generic regardless of lattice structure and dimensionality, but to have explicit examples in mind, we will mainly consider the 2D square and triangular lattices. We focus on the behavior of the model at temperature $T = 0$, although we also make estimates of the parametric dependence of the critical temperatures~\footnote{To extend the effective theories to non-zero temperature $T$, we need to estimate the $T$-scale above which their $T$-dependence is significant. This is especially interesting when $\omega_D$ is small compared to $t$. To make this estimation, we perform perturbation theory averaged on the thermal equilibrium state of the phonons. We find that the values of the coefficients do not significantly change if $U_\text{e-ph},|U_\text{eff}| \gg  \omega_D \epsilon_\beta$ or $U_\text{e-ph},|U_\text{eff}|   \ll \omega_D/\epsilon_\beta$, where $\epsilon_\beta \equiv \frac{2\mathrm{e}^{-\beta \omega_D}}{1-\mathrm{e}^{-\beta \omega_D}}$. This is always satisfied for $T \ll \omega_D$.}. 
Without loss of generality, we will consider the case in which the average number of electrons per site, $n\equiv N^{-1} \sum^N_j\langle \hat n_j\rangle \leq 1$, and will refer to $x=1-n$ as the ``concentration of doped holes.''  (A particle-hole transformation $\hat{c}\leftrightarrow \hat{c}^\dagger$ relates this problem to an electron doped problem with $n=1+x$ and with opposite sign of hopping $t$ and e-ph coupling $\alpha$.) Explicit calculations are deferred to the Supplemental Material~\footnote{See Supplemental Material below, and Refs.~\cite{PhysRevB.48.6141,PhysRevLett.82.3899,Reger_1989,PhysRevLett.104.187203} therein, for detailed discussions of the generalized Holstein-Lang-Firsov transformation, the calculation of various coefficients, and singlet pairing on the triangular lattice.}.

\underline{\bf For  $U_{\text{eff}}>0$},
the ground-state manifold to zeroth order in $t$ consists of all states with no doubly occupied sites and no phonons.  Performing degenerate perturbation theory up to second order yields an effective Hamiltonian %similar to the $t$-$J$-$V$ {\color{red} remove?} model 
(leaving implicit projection onto the space of no doubly-occupied sites and Hermitian conjugation of quantum hopping terms):
\begin{align}\label{t-J-V}
\hat{H}_{\text{eff}} =& - t_1 \sum_{\langle i,j\rangle,\sigma} \hat{c
}_{i,\sigma}^{\dagger}  \hat{c}_{j,\sigma}   -   t_2 \sum_{\langle i, m, j\rangle,\sigma}  \hat{c}_{i,\sigma}^{\dagger} (1-2\hat n_m) \hat{c}_{j,\sigma} \nonumber \\
& - (\tau+2t_2)\sum_{\langle i, m, j\rangle} \hat{s}^{\dagger}_{im} \hat{s}_{mj} 
\nonumber \\
& \  + J\sum_{\langle i,j\rangle }\left[\Vec{{S}}_i\cdot\Vec{{S}}_j - \frac{\hat{n}_i \hat{n}_j}{4} \right] + V \sum_{\langle i,j\rangle} \hat{n}_{i} \hat{n}_{j}
\end{align}
where $t_1$ is the (renormalized) nearest-neighbor hopping, $t_2$ is a next-nearest-neighbor hopping term via an intermediate site $m$ and $\langle i,m,j\rangle$ represents a triplet of sites such that $m$ is a nearest-neighbor of two distinct sites $i$ and $j$, $(\tau+2t_2)$ is a singlet hopping term where $\hat{s}_{ij} = (\hat{c}_{i,\uparrow} \hat{c}_{j,\downarrow}+\hat{c}_{j,\uparrow} \hat{c}_{i,\downarrow})/\sqrt{2}$ is the annihilation operator of a singlet Cooper pair on bond $\langle ij\rangle$, $J$ is the anti-ferromagnetic exchange interaction, and $V$ is the repulsion between electrons on nearest-neighbor sites.  

\begin{figure}[t]
    \centering
    \includegraphics[width=\linewidth]{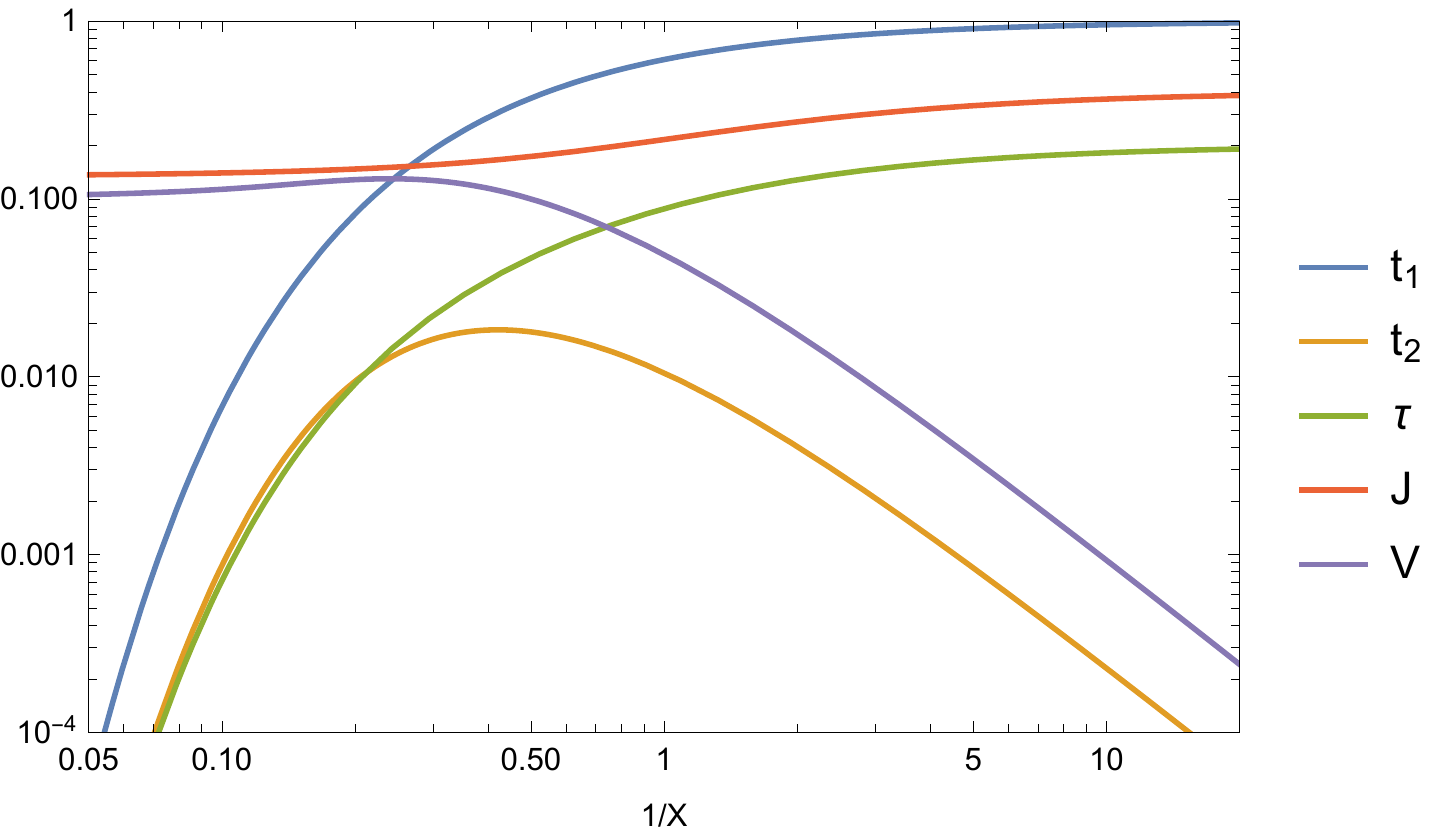}
    \caption{An illustration of the coefficients as functions of the degree of retardation $1/X= \left({\omega_D}/{U_{\text{e-ph}}}\right)$, with fixed $U_{\text{e-e}} = 30$, $U_{\text{e-ph}} = 20$, and $t=1$.}
    \label{illu}
\end{figure}

The values of these effective couplings can be computed explicitly in terms of the dimensionless functions, 
\begin{align}
F(x,y) &\equiv y \mathrm{e}^{-|x|} \int_0^{\infty} \mathrm{d} t \cdot \mathrm{e}^{- y t + x\mathrm{e}^{-t}}  \\
F'(x) 
&\equiv x   \frac{\partial F}{\partial y}\bigg{|}_{y\rightarrow 0} = x\mathrm{e}^{-|x|} \int_0^{\infty} \mathrm{d} t \cdot \left(\mathrm{e}^{ x\mathrm{e}^{-t}}-1\right)
\end{align} 
of the dimensionless  parameters,  $X\equiv \frac{U_{\text{e-ph}}}{\omega_D}$ and $Y\equiv \frac{|U_{\text{eff}}|}{\omega_D}$ as shown  in the first column of Table~\ref{coefficients}, where $\omega_D=\sqrt{k/m}$ is the optical phonon frequency. Explicit asymptotic expressions for these functions can be obtained in the large and small $\omega_D$ limit, as listed in the second and third columns of the table. In Fig.~\ref{illu} we show the coefficients in Eq.~(\ref{t-J-V}) as functions of $X$ for given values of $U_\text{e-e}$, $ U_\text{e-ph}$ and $t$. Increasing the e-ph coupling or lowering the phonon frequency suppresses quantum hopping and thus any tendency toward superconductivity. This suppression is a manifestation of the self-trapping crossover of the single polaron problem. Increasing e-ph coupling also enhances the spin fluctuations, which is consistent with a previous study~\cite{PhysRevLett.97.056402}.

In the anti-adiabatic limit $\omega_D \rightarrow \infty$, the e-e and e-ph interactions  are simply additive, so the effective theory is identical to the standard $t$-$J$ model generated by a Hubbard model with $U=U_{\text{eff}}>0$. In this limit, $|t_1| \gg J$ and $ V$ as usually considered. This hierarchy remains valid in a range of smaller $\omega_D$. While this limit is interesting, and has been widely studied, there is no qualitatively new  physics associated with the presence of phonons.

\begin{table}[t] 
    \centering
\begin{tabular}{ c | c | c }
   & $\omega_D \rightarrow 0$ & $\omega_D \rightarrow \infty$\\
  & ($X,Y\to \infty$)  & ($X, Y\to 0$) \\
  \hline
  \hline
 $t_1=t \mathrm{e}^{-\frac{X}{2}}$   
 & $t \mathrm{e}^{-\frac{X}{2}}$ & $t$ \\
 \hline
 $t_2 = \frac{2t^2}{U_{\text{e-ph}}}  \mathrm{e}^{-\frac{X}{2}} F'(\frac{X}{2})$ &  $\frac{2t^2}{U_{\text{e-ph}}}\mathrm{e}^{-\frac{X}{2}}$ & 
 $\frac{t^2 X^2}{2U_{\text{e-ph}}}$\\ 
 \hline
  $\tau = \frac{2t^2}{U_{\text{eff}}} \mathrm{e}^{-\frac{X}{2}} F(\frac{X}{2},Y)$  & $\frac{4t^2}{U_{\text{eff}}+U_{\text{e-e}}}\mathrm{e}^{-\frac{X}{2}}$ & $\frac{2t^2}{U_{\text{eff}}}$\\ 
  \hline
 $J = \frac{4t^2}{U_{\text{eff}}} F(X,Y) $  & $\frac{4t^2}{U_{\text{e-e}}}$ & $\frac{4t^2}{U_{\text{eff}}}$ \\
 \hline
 $V = \frac{2t^2}{U_{\text{e-ph}}} F'(X)$  &  $\frac{2t^2}{U_{\text{e-ph}}}$ & 
 $\frac{2t^2 X^2}{U_{\text{e-ph}}} $  \\ \hline
\end{tabular}
    \caption{The expressions and the limiting behaviors of the coefficients in the effective theory in Eq.~(\ref{t-J-V}).}
    \label{coefficients}
\end{table}

\begin{figure}
  \centering
  \includegraphics[width=8cm]{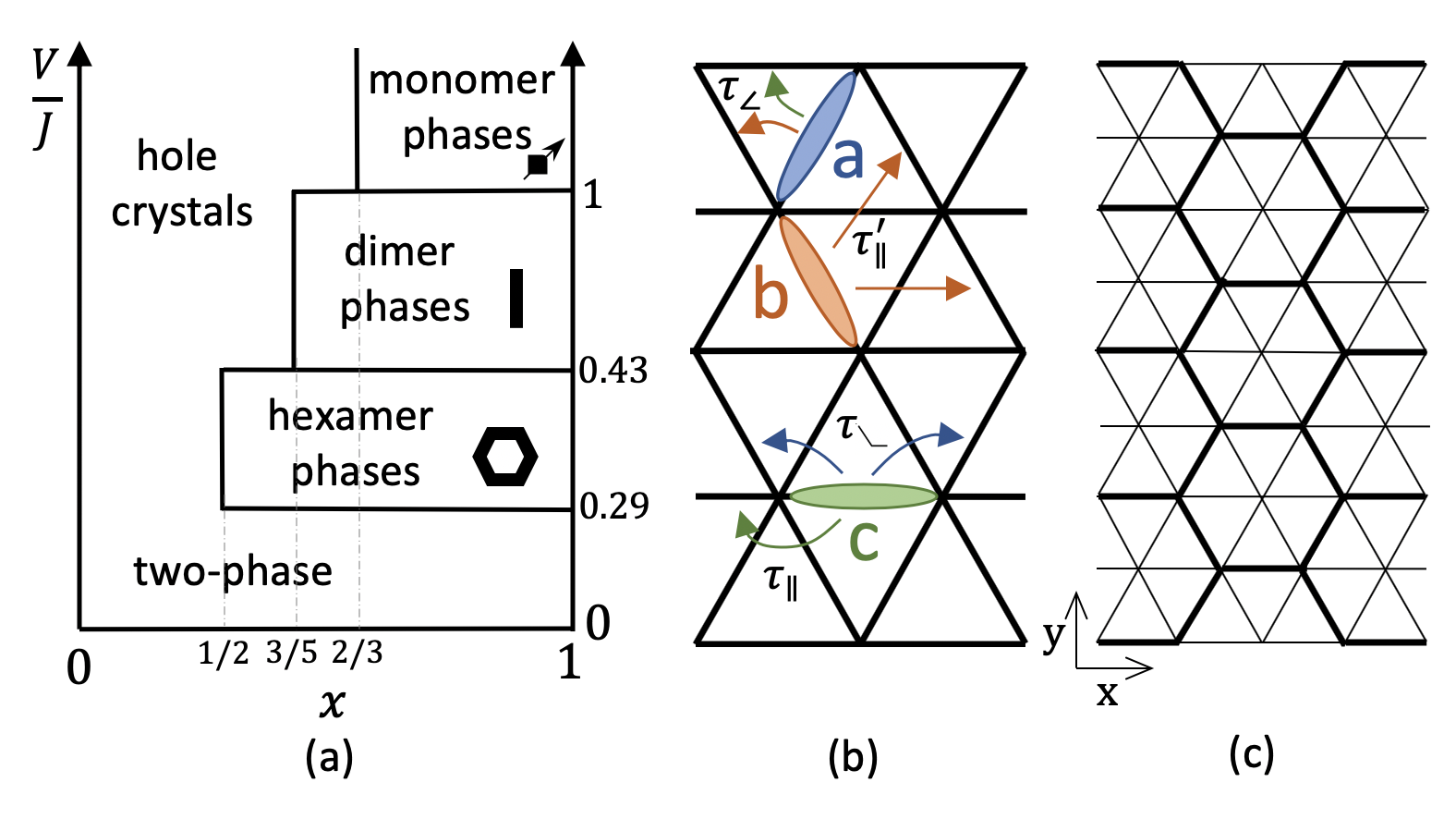}
  \caption{For triangular lattice, (a) a schematic phase diagram for $t_1\ll J,V$ model, (b) an illustration of various pair hopping terms appearing in $\hat{H}_\mathrm{dimer}$ (Eq.~(\ref{dimer})) and (c) a possible PDW pattern, the thickened bonds have pair-field proportional to $+\phi$ and others $-\frac{\phi}{2}$, where $\phi$ is the amplitude.}\label{triangularmain}
\end{figure}

\begin{figure}[t]
    \centering
    \includegraphics[width=\linewidth]{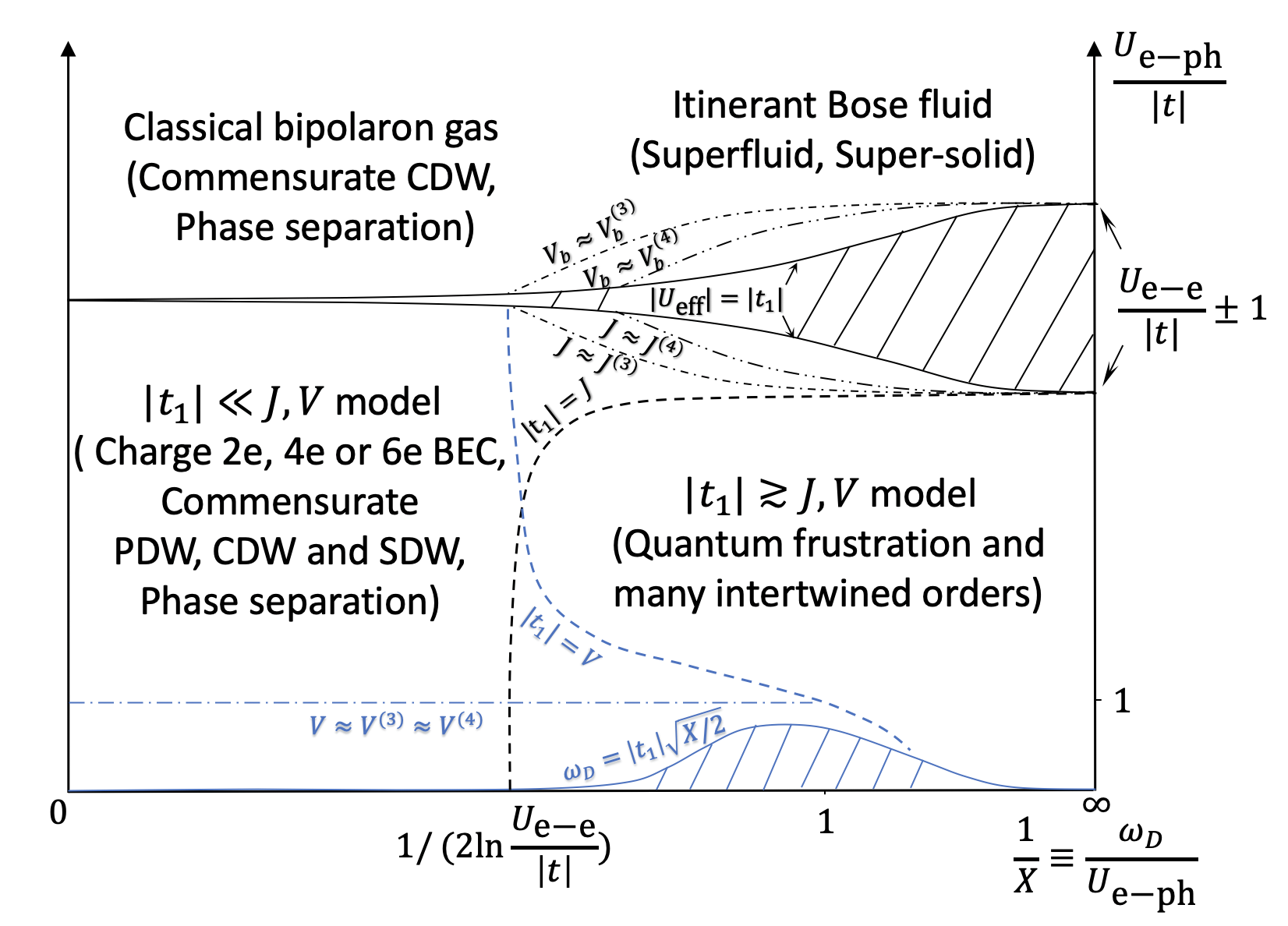}
    \caption{A schematic phase diagram of the Holstein-Hubbard model in the strong-coupling limit. The black or blue dashed lines separate large and small $|t_1|/J$ or $|t_1|/V$ regimes.  The meanings of other lines are discussed in the text, and correspond to the indicated equalities. The condition $V = V^{(m)}$ is not plotted to region $V<|t_1|$, since all orders of $V^{(m)}$ are small compared to quantum hopping in this region.  }
    \label{pd}
\end{figure}

As the phonon frequency is lowered, $J$ and $V$ approach constants, but quantum hoppings are rapidly suppressed, reflecting the effect of a Frank-Condon overlap factor~\cite{lax1952franck}. In the adiabatic limit $\omega_D \rightarrow 0$, the effective model is realized in the limit $|t_1| \ll J,V$, which was previously considered to be unphysical. The effective model is now similar to the small $t$ limit of the $t$-$J$-$V$ model studied in Ref.~\cite{PhysRevB.42.6523}, with the difference that there can also be other smaller hopping terms, $t_2$ and $\tau$. A schematic diagram of possible phases that arise in this limit on the triangular lattice can be seen in Fig.~\ref{triangularmain}(a);  a similar phase diagram was discussed for the square lattice in Fig.~3 of Ref. \cite{PhysRevB.42.6523}. The nature of the resulting phases can be seen by  neglecting the hopping terms to zeroth order.

{\bf Insulating charge and spin density wave phases:} $V/J\ll 1$ leads to two-phase phase coexistence of an insulating antiferromagnet with $x=0$ and an electron-free void with $x=1$, i.e. complete phase separation of the doped holes. At larger values of $V/J$ the phase diagram is more complicated.  Generally, for most values of $x$ smaller than a critical value $x_c$ (which depends on both the lattice geometry and the value of $V/J$), the doped holes form various forms of commensurate hole crystals coexisting with some form of antiferromagnetic order, likely forming some form of two-phase coexistence between two such phases. An example of this is the $\sqrt{5}\times\sqrt{5}$ hole crystal with $x=1/5$ discussed for the square lattice in Ref.~\cite{PhysRevB.42.6523}, and an analogous $\sqrt{7}\times \sqrt{7}$ hole crystal that likely arises on the triangular lattice with $x=1/7$.  

A variety of more unusual behaviors arise at lower electron density.
When $x> x_c$, the system can be thought of as a dilute collection of electrons, which form small disconnected clusters. The effect of the relatively smaller hopping terms then resolves remaining ground-state degeneracy by degenerate perturbation theory.

{\bf Heavy Fermi liquid:} For large $V/J>1$, monomers are favored to the zeroth order. Extensively degenerate ground-states consist of all configuration wheres no pair of nearest-neighbor sites is occupied. When the hopping terms are included, these monomers can be treated as  spin-$1/2$ fermions with a hard-core radius that extends to nearest-neighbor sites, and a highly renormalized hopping matrix element, $t_1$.  Typically we would expect this system to form a heavy Fermi liquid, although at various commensurate values of $x$, via ``order by disorder,'' it may well exhibit commensurate CDW order with some form of accompanying SDW order~\cite{doi:10.1146/annurev-conmatphys-033117-053925}. It also can have a very low $T$ Kohn-Luttinger type instability to unconventional superconductivity~\cite{PhysRevLett.15.524}.

{\bf Hard Core Dimer fluid:}  For $\nu_2< V/J < 1$ (where $\nu_2=0.5$ and $0.43$ for the square and triangular lattices respectively), singlet pairs of electrons on nearest neighbor bonds are energetically optimal. These dimers are eliptical hard-core bosons~\footnote{In the context of bipolaron problem, extended versions of this have been called S1 bipolarons~\cite{PROVILLE1998307, PhysRevB.69.245111}.}, and the zeroth order ground states can be labeled by dimer configurations where the dimers satisfy both a hard-core constraint (no two dimers touch the same site) and a nearest-neighbor exclusion (no pair of nearest-neighbor sites is touched by distinct dimers).

The ground-state degeneracy is lifted when the effect of  hopping terms is included.  While at special commensurate densities, this could lead to an insulating CDW phase, generically it leads to charge $2e$ singlet superfluid phases of various sorts. To address the nature of these phases, we write the effective model of hard-core dimers:
\begin{eqnarray}
\hat{H}_\mathrm{dimer}=-\sum_{\langle ij\rangle,\langle mn\rangle} \left( \tau_{ij,mn} \hat s^\dag_{ij} \hat s_{mn} +\textrm{h.c.}\right),
\label{dimer}
\end{eqnarray}
where $\tau_{ij,mn}$ is the effective pair hopping amplitude between bond $\langle ij\rangle$ and bond $\langle mn\rangle$. There are various distinct types of hopping processes that can arise up to second order in $t$ and contribute to different sorts of dimer hopping amplitudes. Generically, dimers can be moved by the singlet-pair hopping $(\tau+2t_2)$ and next-nearest-neighbor hopping $t_2$. (Another virtual process with amplitude $\frac{t_1^2}{J-V}$ can also hop dimers, but it is unimportant in the adiabatic limit since it is suppressed relative to the terms we have kept by a factor of $\mathrm{e}^{-\frac{X}{2}}$.) These processes in the second order of $t$ all make positive contributions to $\tau_{ij,mn}$, independent of the sign of $t$.  When all $\tau_{ij,mn}\geq 0$, $-\hat{H}_\mathrm{dimer}$ satisfies the conditions of Perron-Frobenius theorem~\cite{maccluer2000many}, and hence the Hamiltonian is minimized by a Bloch state with $\vec k=\vec 0$ and all positive amplitudes. Since at finite dimer density, the system likely forms a Bose condensate, this implies the pair-field is spatially uniform on square and most lattices.

However, on triangular lattice (or other frustrated geometries), dimers can hop via a first order process in $t_1$. This process contributes to the type of pair hopping labeled $\tau_{\angle{}}$ in Fig.~\ref{triangularmain}(b), in which one end of a dimer pivots by $60^\circ$ about the other end, and it has much larger amplitude than $\tau$ and $t_2$. Consequently, $\tau_{\angle{}}$ has the same sign as of $t$, and is larger in magnitude than the remaining terms $\tau_\parallel$, $\tau'_\parallel$, and $\tau_{\obtuseangle{}}$. This opens the possibility of exotic condensation when $t<0$. On a triangular lattice, there are three possible dimer states per unit cell, and correspondingly three bands.  Taking into account only the largest pair-hopping term, $\tau_{\angle{}}$, these consist of a flat band and two dispersing bands, such that the flat band is the lowest if $t<0$. Including the effects of the smaller pair-hopping terms, we find the band minima occur at the K and $-$K points in the Brillouin Zone. A dimer Bose-condensate thus results in some form of a PDW~\cite{agterberg2020physics}.  A state in which the condensed bosons have momentum either K or $-$K  breaks time-reversal symmetry, but has a spatially uniform magnitude of pair-field. If time-reversal symmetry is preserved, singlets equally condense in $\pm$K, resulting in a translation symmetry breaking pattern of the pair-field, as shown in Fig.~\ref{triangularmain}(c). {Which form of condensate is favored remains to be determined due to the strongly interacting nature of bosons in the present problem although it was shown that the plane-wave state is favored for the case of weakly interacting bosons~\cite{HuiZhai_2012}.} For a Bose condensate at either $\Gamma$ or K point, the curvature of the band bottom is set by terms to the second order in $t$ and  the superconducting transition temperature is parametrically small, $T_c\sim t^2 \mathrm{e}^{-\frac{X}{2}}$, in the adiabatic limit.

{\bf Polygonal fluid:} Further reducing $V/J$, different lattices lead to different optimal clusters. On a  square lattice,  a tetramer (square) minimizes the energy for $\nu_4< V/J < \nu_2 = 0.5$, and phase separation occurs for $V/J < \nu_4=0.418$. On a triangular/honeycomb/Kagome lattice, a hexamer minimizes the energy for $\nu_6< V/J<\nu_2 = 0.434$, and phase separation occurs for $V/J <\nu_6 = 0.290$/$0.390$/$0.215$. These clusters are hard-core bosons with nearest-neighbor exclusion that can condense into charge $4e$ or $6e$ superconducting phases. The quantum hopping of these clusters derives from high order process, so the superconducting transition temperature should be low and upper-bounded by $t_1^p$, where $p$ is the number of electrons in the cluster.

It is not always the case that the simple Holstein-Hubbard model can realize all the interesting ranges of $V/J$.  In particular, in the adiabatic limit, $V/J = \frac{U_{\text{e-e}}}{2U_{\text{e-ph}}}>0.5$. However, the introduction of weak dispersion of the phonons or longer-ranged coupling to the electron densities will introduce nearest-neighbor attraction in the zeroth order that can be easily comparable with $J$, and drive the system into the interesting cluster-phases. As a concrete example, adding weak phonon coupling to nearest neighbor sites $\hat{H}' = \sum_{<i,j>} \alpha' (\hat{n}_i + \hat{n}_j) (\hat{x}_i + \hat{x}_j)$ introduces a small nearest-neighbor attraction $\frac{\alpha'^2}{k}$ without modifying the hopping terms.

\underline{\bf For $U_{\text{eff}}<0$},  the degenerate ground state manifold in the absence of hopping consists of states occupied by pairs of electrons (on-site bipolarons) and no phonon. Including hopping in degenerate perturbation theory yields
\begin{align}\label{bipolaron}
\hat{H}_{\text{eff}} = - &t_b \sum_{\langle i,j\rangle} (\hat{b
}_{i}^{\dagger} \hat{b}_{j} + \text{h.c.}) + V_b \sum_{\langle i,j\rangle} \hat{b
}_{i}^{\dagger} \hat{b}_{i} \hat{b
}_{j}^{\dagger} \hat{b}_{j}
\end{align}
where $\hat{b}_i \equiv \hat{c}_{i,\uparrow} \hat{c}_{i,\downarrow}$ annihilates a hard-core boson on site-$i$ ($\hat b_i^\dagger \hat b_i = 0, 1$ is implicitly imposed). $t_b$ and $V_b$  are, respectively, the nearest-neighbor hopping and repulsion whose values are listed in Table~\ref{bicoefficients}. 
\begin{table}[t] 
    \centering
\begin{tabular}{ c | c | c }
   & $\omega_D \rightarrow 0$ & $\omega_D \rightarrow \infty$ \\
  \hline
  \hline
 $t_b=\frac{2t^2}{|U_{\text{eff}}|} F(-X,Y)$ & $\frac{2t^2}{|U_{\text{eff}}|}\frac{U_\text{e-ph}}{U_\text{e-e}}\mathrm{e}^{-(X+Y)}$ & $\frac{2t^2}{|U_{\text{eff}}|}$ \\ 
 \hline
 $V_b = \frac{4t^2}{|U_{\text{eff}}|} F(X,Y)$  & $\frac{4t^2}{U_{\text{e-ph}}+|U_{\text{eff}}|}$ &  $\frac{4t^2}{|U_{\text{eff}}|}$\\
 \hline
\end{tabular}
     \caption{The expressions and the limiting behaviors of the coefficients in the effective theory in Eq.~(\ref{bipolaron}).}
    \label{bicoefficients}
\end{table}
This is a standard hard-core boson model (or equivalently a spin-$1/2$ XXZ model), for which superfluidity and charge orders were investigated on various lattices~\cite{PhysRevLett.88.167208,PhysRevB.77.014524,PhysRevLett.95.127207,PhysRevLett.102.017203,PhysRevB.75.174301}.  Below a Kosterlitz-Thouless transition $\sim t_b$, superfluidity is possible. This transition temperature is also parameterically small, $T_c \lesssim (t^2/|U_\mathrm{eff}|)\mathrm{e}^{-X}$. Particular interesting possibilities are supersolid phases on frustrated lattices, where the predicted phase region $t_b \leq 0.1 V_b$ for triangular lattice~\cite{PhysRevLett.102.017203} is clearly accessible through tuning retardation. Indeed, coexisting superconducting and charge orders were predicted theoretically~\cite{Stein_1997} and have been seen in a recent study of the triangular lattice Holstein model~\cite{PhysRevB.100.245105}. Above the superfluid transition temperature, but below the binding energy $|U_\text{eff}|$, the system is essentially a classical bipolaron gas, where various commensurate charge orders and phase separations can exist below an Ising critical temperature $\sim V_b$~\cite{PhysRevB.98.085405}. Increasing the e-e repulsion can enhance charge and especially superconducting order, in contrast to a previous study on the weak coupling regime~\cite{PhysRevLett.75.2570}.

\underline{\bf Range of validity of the effective theories:} The effective models we have derived operate in reduced Hilbert spaces with restricted site occupancies (determined by the sign of $U_{\text{eff}}$) and zero phonon excitations. These restrictions become invalid when excitation energies in the unperturbed Hamiltonian are no longer large compared with the corresponding perturbation matrix elements. Specifically, all possible site occupancies should be considered when $|U_{\text{eff}}|\lesssim|t_1|=|t|\mathrm{e}^{-X/2}$, and phonon excitations should be included if $\omega_D\lesssim|t_1|\sqrt{\frac{X}{2}}$. These narrow regions are enclosed with solid lines and hashed out in the schematic phase diagram in Fig.~\ref{pd}. 

Within the reduced Hilbert space, there remains the issue of whether it is sufficient to compute the effective interactions to low order in powers of $|t|$. This is controlled so long as the longer-ranged interactions generated by higher order terms  are small compared with the terms we have already considered. This sort of analysis was carried out for the strong-coupling limit of Holstein and Hubbard models in Refs.~\cite{PhysRevB.37.9753,PhysRevB.48.3881}. For the $m$-th order of $J^{(m)}$, $V^{(m)}$ or $V_b^{(m)}$ series, we evaluate the amplitudes of virtual processes involving hopping around $m$ sites, and regard them as representative. The condition (valid so long as  $U_\text{e-e} \gg |U_\text{eff}| \sim |t|$) for $J \gg J^{(m)}$ and $V_b \gg V_b^{(m)}$ is: $|U_\text{eff}| \gg \text{min}\{|t|, r_m  U_\text{e-ph} \mathrm{e}^{-X}\}$, where $r_m \equiv \frac{m}{2(m-1)}$. Similarly, $V \gg V^{(m)}$ so long as $U_\text{e-ph} \gg |t|\cdot \text{min}\{1,X\cdot2^{-\frac{1}{m-2}}\}$.
The black and blue dashed-dotted lines in Fig.~\ref{pd} are thus defined, as indicated, by the estimation $J\approx J^{(m)}$, $V_b\approx V_b^{(m)}$ or $V \approx V^{(m)}$ for $m=3$ (for triangular lattice) and $m=4$ (for square lattice). Longer-ranged interactions are significant inside these lines.

{\bf Acknowledgements:}  S.A.K. was supported, in part, by NSF grant No. DMR-2000987 at Stanford. H.Y.was supported in part by National Science Fund for Distinguished Young Scholars of NSFC through Grant No. 11825404 at Tsinghua, by the Ministry of Science and Technology of China under Grant No. 2016YFA0301001 and 2018YFA0305604, as well as by the Gordon and Betty Moore Foundation’s EPiQS through Grant No. GBMF4302 at Stanford. 

\bibliographystyle{apsrev4-1} 
\bibliography{Holstein}

\begin{thebibliography}{51}%
\makeatletter
\providecommand \@ifxundefined [1]{%
 \@ifx{#1\undefined}
}%
\providecommand \@ifnum [1]{%
 \ifnum #1\expandafter \@firstoftwo
 \else \expandafter \@secondoftwo
 \fi
}%
\providecommand \@ifx [1]{%
 \ifx #1\expandafter \@firstoftwo
 \else \expandafter \@secondoftwo
 \fi
}%
\providecommand \natexlab [1]{#1}%
\providecommand \enquote  [1]{``#1''}%
\providecommand \bibnamefont  [1]{#1}%
\providecommand \bibfnamefont [1]{#1}%
\providecommand \citenamefont [1]{#1}%
\providecommand \href@noop [0]{\@secondoftwo}%
\providecommand \href [0]{\begingroup \@sanitize@url \@href}%
\providecommand \@href[1]{\@@startlink{#1}\@@href}%
\providecommand \@@href[1]{\endgroup#1\@@endlink}%
\providecommand \@sanitize@url [0]{\catcode `\\12\catcode `\$12\catcode
  `\&12\catcode `\#12\catcode `\^12\catcode `\_12\catcode `\%12\relax}%
\providecommand \@@startlink[1]{}%
\providecommand \@@endlink[0]{}%
\providecommand \url  [0]{\begingroup\@sanitize@url \@url }%
\providecommand \@url [1]{\endgroup\@href {#1}{\urlprefix }}%
\providecommand \urlprefix  [0]{URL }%
\providecommand \Eprint [0]{\href }%
\providecommand \doibase [0]{http://dx.doi.org/}%
\providecommand \selectlanguage [0]{\@gobble}%
\providecommand \bibinfo  [0]{\@secondoftwo}%
\providecommand \bibfield  [0]{\@secondoftwo}%
\providecommand \translation [1]{[#1]}%
\providecommand \BibitemOpen [0]{}%
\providecommand \bibitemStop [0]{}%
\providecommand \bibitemNoStop [0]{.\EOS\space}%
\providecommand \EOS [0]{\spacefactor3000\relax}%
\providecommand \BibitemShut  [1]{\csname bibitem#1\endcsname}%
\let\auto@bib@innerbib\@empty
%</preamble>
\bibitem [{\citenamefont {Zhong}\ and\ \citenamefont
  {Sch\"uttler}(1992)}]{PhysRevLett.69.1600}%
  \BibitemOpen
  \bibfield  {author} {\bibinfo {author} {\bibfnamefont {J.}~\bibnamefont
  {Zhong}}\ and\ \bibinfo {author} {\bibfnamefont {H.-B.}\ \bibnamefont
  {Sch\"uttler}},\ }\href {\doibase 10.1103/PhysRevLett.69.1600} {\bibfield
  {journal} {\bibinfo  {journal} {Phys. Rev. Lett.}\ }\textbf {\bibinfo
  {volume} {69}},\ \bibinfo {pages} {1600} (\bibinfo {year}
  {1992})}\BibitemShut {NoStop}%
\bibitem [{\citenamefont {Freericks}(1993)}]{PhysRevB.48.3881}%
  \BibitemOpen
  \bibfield  {author} {\bibinfo {author} {\bibfnamefont {J.~K.}\ \bibnamefont
  {Freericks}},\ }\href {\doibase 10.1103/PhysRevB.48.3881} {\bibfield
  {journal} {\bibinfo  {journal} {Phys. Rev. B}\ }\textbf {\bibinfo {volume}
  {48}},\ \bibinfo {pages} {3881} (\bibinfo {year} {1993})}\BibitemShut
  {NoStop}%
\bibitem [{\citenamefont {Freericks}\ and\ \citenamefont
  {Jarrell}(1994)}]{PhysRevB.50.6939}%
  \BibitemOpen
  \bibfield  {author} {\bibinfo {author} {\bibfnamefont {J.~K.}\ \bibnamefont
  {Freericks}}\ and\ \bibinfo {author} {\bibfnamefont {M.}~\bibnamefont
  {Jarrell}},\ }\href {\doibase 10.1103/PhysRevB.50.6939} {\bibfield  {journal}
  {\bibinfo  {journal} {Phys. Rev. B}\ }\textbf {\bibinfo {volume} {50}},\
  \bibinfo {pages} {6939} (\bibinfo {year} {1994})}\BibitemShut {NoStop}%
\bibitem [{\citenamefont {Fehske}\ \emph {et~al.}(1994)\citenamefont {Fehske},
  \citenamefont {Ihle}, \citenamefont {Loos}, \citenamefont {Trapper},\ and\
  \citenamefont {B{\"u}ttner}}]{fehske1994polaron}%
  \BibitemOpen
  \bibfield  {author} {\bibinfo {author} {\bibfnamefont {H.}~\bibnamefont
  {Fehske}}, \bibinfo {author} {\bibfnamefont {D.}~\bibnamefont {Ihle}},
  \bibinfo {author} {\bibfnamefont {J.}~\bibnamefont {Loos}}, \bibinfo {author}
  {\bibfnamefont {U.}~\bibnamefont {Trapper}}, \ and\ \bibinfo {author}
  {\bibfnamefont {H.}~\bibnamefont {B{\"u}ttner}},\ }\href@noop {} {\bibfield
  {journal} {\bibinfo  {journal} {Zeitschrift f{\"u}r Physik B Condensed
  Matter}\ }\textbf {\bibinfo {volume} {94}},\ \bibinfo {pages} {91} (\bibinfo
  {year} {1994})}\BibitemShut {NoStop}%
\bibitem [{\citenamefont {Berger}\ \emph {et~al.}(1995)\citenamefont {Berger},
  \citenamefont {Val\'a\ifmmode~\check{s}\else \v{s}\fi{}ek},\ and\
  \citenamefont {von~der Linden}}]{PhysRevB.52.4806}%
  \BibitemOpen
  \bibfield  {author} {\bibinfo {author} {\bibfnamefont {E.}~\bibnamefont
  {Berger}}, \bibinfo {author} {\bibfnamefont {P.}~\bibnamefont
  {Val\'a\ifmmode~\check{s}\else \v{s}\fi{}ek}}, \ and\ \bibinfo {author}
  {\bibfnamefont {W.}~\bibnamefont {von~der Linden}},\ }\href {\doibase
  10.1103/PhysRevB.52.4806} {\bibfield  {journal} {\bibinfo  {journal} {Phys.
  Rev. B}\ }\textbf {\bibinfo {volume} {52}},\ \bibinfo {pages} {4806}
  (\bibinfo {year} {1995})}\BibitemShut {NoStop}%
\bibitem [{\citenamefont {Freericks}\ and\ \citenamefont
  {Jarrell}(1995)}]{PhysRevLett.75.2570}%
  \BibitemOpen
  \bibfield  {author} {\bibinfo {author} {\bibfnamefont {J.~K.}\ \bibnamefont
  {Freericks}}\ and\ \bibinfo {author} {\bibfnamefont {M.}~\bibnamefont
  {Jarrell}},\ }\href {\doibase 10.1103/PhysRevLett.75.2570} {\bibfield
  {journal} {\bibinfo  {journal} {Phys. Rev. Lett.}\ }\textbf {\bibinfo
  {volume} {75}},\ \bibinfo {pages} {2570} (\bibinfo {year}
  {1995})}\BibitemShut {NoStop}%
\bibitem [{\citenamefont {Wellein}\ \emph {et~al.}(1996)\citenamefont
  {Wellein}, \citenamefont {R\"oder},\ and\ \citenamefont
  {Fehske}}]{PhysRevB.53.9666}%
  \BibitemOpen
  \bibfield  {author} {\bibinfo {author} {\bibfnamefont {G.}~\bibnamefont
  {Wellein}}, \bibinfo {author} {\bibfnamefont {H.}~\bibnamefont {R\"oder}}, \
  and\ \bibinfo {author} {\bibfnamefont {H.}~\bibnamefont {Fehske}},\ }\href
  {\doibase 10.1103/PhysRevB.53.9666} {\bibfield  {journal} {\bibinfo
  {journal} {Phys. Rev. B}\ }\textbf {\bibinfo {volume} {53}},\ \bibinfo
  {pages} {9666} (\bibinfo {year} {1996})}\BibitemShut {NoStop}%
\bibitem [{\citenamefont {Stein}(1997)}]{Stein_1997}%
  \BibitemOpen
  \bibfield  {author} {\bibinfo {author} {\bibfnamefont {J.}~\bibnamefont
  {Stein}},\ }\href {\doibase 10.1209/epl/i1997-00370-1} {\bibfield  {journal}
  {\bibinfo  {journal} {Europhysics Letters ({EPL})}\ }\textbf {\bibinfo
  {volume} {39}},\ \bibinfo {pages} {413} (\bibinfo {year} {1997})}\BibitemShut
  {NoStop}%
\bibitem [{\citenamefont {Proville}\ and\ \citenamefont
  {Aubry}(1998)}]{PROVILLE1998307}%
  \BibitemOpen
  \bibfield  {author} {\bibinfo {author} {\bibfnamefont {L.}~\bibnamefont
  {Proville}}\ and\ \bibinfo {author} {\bibfnamefont {S.}~\bibnamefont
  {Aubry}},\ }\href {\doibase https://doi.org/10.1016/S0167-2789(97)00283-2}
  {\bibfield  {journal} {\bibinfo  {journal} {Physica D: Nonlinear Phenomena}\
  }\textbf {\bibinfo {volume} {113}},\ \bibinfo {pages} {307 } (\bibinfo {year}
  {1998})},\ \bibinfo {note} {proceedings of the Conference on Fluctuations,
  Nonlinearity and Disorder in Condensed Matter and Biological
  Physics}\BibitemShut {NoStop}%
\bibitem [{\citenamefont {Bon\ifmmode~\check{c}\else \v{c}\fi{}a}\ and\
  \citenamefont {Trugman}(2001)}]{PhysRevB.64.094507}%
  \BibitemOpen
  \bibfield  {author} {\bibinfo {author} {\bibfnamefont {J.}~\bibnamefont
  {Bon\ifmmode~\check{c}\else \v{c}\fi{}a}}\ and\ \bibinfo {author}
  {\bibfnamefont {S.~A.}\ \bibnamefont {Trugman}},\ }\href {\doibase
  10.1103/PhysRevB.64.094507} {\bibfield  {journal} {\bibinfo  {journal} {Phys.
  Rev. B}\ }\textbf {\bibinfo {volume} {64}},\ \bibinfo {pages} {094507}
  (\bibinfo {year} {2001})}\BibitemShut {NoStop}%
\bibitem [{\citenamefont {Takada}\ and\ \citenamefont
  {Chatterjee}(2003)}]{PhysRevB.67.081102}%
  \BibitemOpen
  \bibfield  {author} {\bibinfo {author} {\bibfnamefont {Y.}~\bibnamefont
  {Takada}}\ and\ \bibinfo {author} {\bibfnamefont {A.}~\bibnamefont
  {Chatterjee}},\ }\href {\doibase 10.1103/PhysRevB.67.081102} {\bibfield
  {journal} {\bibinfo  {journal} {Phys. Rev. B}\ }\textbf {\bibinfo {volume}
  {67}},\ \bibinfo {pages} {081102} (\bibinfo {year} {2003})}\BibitemShut
  {NoStop}%
\bibitem [{\citenamefont {Capone}\ \emph {et~al.}(2004)\citenamefont {Capone},
  \citenamefont {Sangiovanni}, \citenamefont {Castellani}, \citenamefont
  {Di~Castro},\ and\ \citenamefont {Grilli}}]{PhysRevLett.92.106401}%
  \BibitemOpen
  \bibfield  {author} {\bibinfo {author} {\bibfnamefont {M.}~\bibnamefont
  {Capone}}, \bibinfo {author} {\bibfnamefont {G.}~\bibnamefont {Sangiovanni}},
  \bibinfo {author} {\bibfnamefont {C.}~\bibnamefont {Castellani}}, \bibinfo
  {author} {\bibfnamefont {C.}~\bibnamefont {Di~Castro}}, \ and\ \bibinfo
  {author} {\bibfnamefont {M.}~\bibnamefont {Grilli}},\ }\href {\doibase
  10.1103/PhysRevLett.92.106401} {\bibfield  {journal} {\bibinfo  {journal}
  {Phys. Rev. Lett.}\ }\textbf {\bibinfo {volume} {92}},\ \bibinfo {pages}
  {106401} (\bibinfo {year} {2004})}\BibitemShut {NoStop}%
\bibitem [{\citenamefont {Macridin}\ \emph {et~al.}(2004)\citenamefont
  {Macridin}, \citenamefont {Sawatzky},\ and\ \citenamefont
  {Jarrell}}]{PhysRevB.69.245111}%
  \BibitemOpen
  \bibfield  {author} {\bibinfo {author} {\bibfnamefont {A.}~\bibnamefont
  {Macridin}}, \bibinfo {author} {\bibfnamefont {G.~A.}\ \bibnamefont
  {Sawatzky}}, \ and\ \bibinfo {author} {\bibfnamefont {M.}~\bibnamefont
  {Jarrell}},\ }\href {\doibase 10.1103/PhysRevB.69.245111} {\bibfield
  {journal} {\bibinfo  {journal} {Phys. Rev. B}\ }\textbf {\bibinfo {volume}
  {69}},\ \bibinfo {pages} {245111} (\bibinfo {year} {2004})}\BibitemShut
  {NoStop}%
\bibitem [{\citenamefont {Koller}\ \emph
  {et~al.}(2004{\natexlab{a}})\citenamefont {Koller}, \citenamefont {Meyer},\
  and\ \citenamefont {Hewson}}]{PhysRevB.70.155103}%
  \BibitemOpen
  \bibfield  {author} {\bibinfo {author} {\bibfnamefont {W.}~\bibnamefont
  {Koller}}, \bibinfo {author} {\bibfnamefont {D.}~\bibnamefont {Meyer}}, \
  and\ \bibinfo {author} {\bibfnamefont {A.~C.}\ \bibnamefont {Hewson}},\
  }\href {\doibase 10.1103/PhysRevB.70.155103} {\bibfield  {journal} {\bibinfo
  {journal} {Phys. Rev. B}\ }\textbf {\bibinfo {volume} {70}},\ \bibinfo
  {pages} {155103} (\bibinfo {year} {2004}{\natexlab{a}})}\BibitemShut
  {NoStop}%
\bibitem [{\citenamefont {Koller}\ \emph
  {et~al.}(2004{\natexlab{b}})\citenamefont {Koller}, \citenamefont {Meyer},
  \citenamefont {{\={O}}no},\ and\ \citenamefont {Hewson}}]{Koller_2004}%
  \BibitemOpen
  \bibfield  {author} {\bibinfo {author} {\bibfnamefont {W.}~\bibnamefont
  {Koller}}, \bibinfo {author} {\bibfnamefont {D.}~\bibnamefont {Meyer}},
  \bibinfo {author} {\bibfnamefont {Y.}~\bibnamefont {{\={O}}no}}, \ and\
  \bibinfo {author} {\bibfnamefont {A.~C.}\ \bibnamefont {Hewson}},\ }\href
  {\doibase 10.1209/epl/i2003-10228-6} {\bibfield  {journal} {\bibinfo
  {journal} {Europhysics Letters ({EPL})}\ }\textbf {\bibinfo {volume} {66}},\
  \bibinfo {pages} {559} (\bibinfo {year} {2004}{\natexlab{b}})}\BibitemShut
  {NoStop}%
\bibitem [{\citenamefont {Carlson}\ \emph {et~al.}(2004)\citenamefont
  {Carlson}, \citenamefont {Kivelson}, \citenamefont {Orgad},\ and\
  \citenamefont {Emery}}]{Carlson2004}%
  \BibitemOpen
  \bibfield  {author} {\bibinfo {author} {\bibfnamefont {E.~W.}\ \bibnamefont
  {Carlson}}, \bibinfo {author} {\bibfnamefont {S.~A.}\ \bibnamefont
  {Kivelson}}, \bibinfo {author} {\bibfnamefont {D.}~\bibnamefont {Orgad}}, \
  and\ \bibinfo {author} {\bibfnamefont {V.~J.}\ \bibnamefont {Emery}},\
  }\enquote {\bibinfo {title} {Concepts in high temperature
  superconductivity},}\ in\ \href {\doibase 10.1007/978-3-642-18914-2_6} {\emph
  {\bibinfo {booktitle} {The Physics of Superconductors: Vol. II.
  Superconductivity in Nanostructures, High-Tc and Novel Superconductors,
  Organic Superconductors}}},\ \bibinfo {editor} {edited by\ \bibinfo {editor}
  {\bibfnamefont {K.~H.}\ \bibnamefont {Bennemann}}\ and\ \bibinfo {editor}
  {\bibfnamefont {J.~B.}\ \bibnamefont {Ketterson}}}\ (\bibinfo  {publisher}
  {Springer Berlin Heidelberg},\ \bibinfo {address} {Berlin, Heidelberg},\
  \bibinfo {year} {2004})\ pp.\ \bibinfo {pages} {275--451}\BibitemShut
  {NoStop}%
\bibitem [{\citenamefont {Macridin}\ \emph {et~al.}(2006)\citenamefont
  {Macridin}, \citenamefont {Moritz}, \citenamefont {Jarrell},\ and\
  \citenamefont {Maier}}]{PhysRevLett.97.056402}%
  \BibitemOpen
  \bibfield  {author} {\bibinfo {author} {\bibfnamefont {A.}~\bibnamefont
  {Macridin}}, \bibinfo {author} {\bibfnamefont {B.}~\bibnamefont {Moritz}},
  \bibinfo {author} {\bibfnamefont {M.}~\bibnamefont {Jarrell}}, \ and\
  \bibinfo {author} {\bibfnamefont {T.}~\bibnamefont {Maier}},\ }\href
  {\doibase 10.1103/PhysRevLett.97.056402} {\bibfield  {journal} {\bibinfo
  {journal} {Phys. Rev. Lett.}\ }\textbf {\bibinfo {volume} {97}},\ \bibinfo
  {pages} {056402} (\bibinfo {year} {2006})}\BibitemShut {NoStop}%
\bibitem [{\citenamefont {Werner}\ and\ \citenamefont
  {Millis}(2007)}]{PhysRevLett.99.146404}%
  \BibitemOpen
  \bibfield  {author} {\bibinfo {author} {\bibfnamefont {P.}~\bibnamefont
  {Werner}}\ and\ \bibinfo {author} {\bibfnamefont {A.~J.}\ \bibnamefont
  {Millis}},\ }\href {\doibase 10.1103/PhysRevLett.99.146404} {\bibfield
  {journal} {\bibinfo  {journal} {Phys. Rev. Lett.}\ }\textbf {\bibinfo
  {volume} {99}},\ \bibinfo {pages} {146404} (\bibinfo {year}
  {2007})}\BibitemShut {NoStop}%
\bibitem [{\citenamefont {Hardikar}\ and\ \citenamefont
  {Clay}(2007)}]{PhysRevB.75.245103}%
  \BibitemOpen
  \bibfield  {author} {\bibinfo {author} {\bibfnamefont {R.~P.}\ \bibnamefont
  {Hardikar}}\ and\ \bibinfo {author} {\bibfnamefont {R.~T.}\ \bibnamefont
  {Clay}},\ }\href {\doibase 10.1103/PhysRevB.75.245103} {\bibfield  {journal}
  {\bibinfo  {journal} {Phys. Rev. B}\ }\textbf {\bibinfo {volume} {75}},\
  \bibinfo {pages} {245103} (\bibinfo {year} {2007})}\BibitemShut {NoStop}%
\bibitem [{\citenamefont {Gunnarsson}\ and\ \citenamefont
  {Rösch}(2008)}]{Gunnarsson_2008}%
  \BibitemOpen
  \bibfield  {author} {\bibinfo {author} {\bibfnamefont {O.}~\bibnamefont
  {Gunnarsson}}\ and\ \bibinfo {author} {\bibfnamefont {O.}~\bibnamefont
  {Rösch}},\ }\href {\doibase 10.1088/0953-8984/20/04/043201} {\bibfield
  {journal} {\bibinfo  {journal} {Journal of Physics: Condensed Matter}\
  }\textbf {\bibinfo {volume} {20}},\ \bibinfo {pages} {043201} (\bibinfo
  {year} {2008})}\BibitemShut {NoStop}%
\bibitem [{\citenamefont {Kumar}\ and\ \citenamefont {van~den
  Brink}(2008)}]{PhysRevB.78.155123}%
  \BibitemOpen
  \bibfield  {author} {\bibinfo {author} {\bibfnamefont {S.}~\bibnamefont
  {Kumar}}\ and\ \bibinfo {author} {\bibfnamefont {J.}~\bibnamefont {van~den
  Brink}},\ }\href {\doibase 10.1103/PhysRevB.78.155123} {\bibfield  {journal}
  {\bibinfo  {journal} {Phys. Rev. B}\ }\textbf {\bibinfo {volume} {78}},\
  \bibinfo {pages} {155123} (\bibinfo {year} {2008})}\BibitemShut {NoStop}%
\bibitem [{\citenamefont {Hohenadler}\ and\ \citenamefont
  {Assaad}(2013)}]{PhysRevB.87.075149}%
  \BibitemOpen
  \bibfield  {author} {\bibinfo {author} {\bibfnamefont {M.}~\bibnamefont
  {Hohenadler}}\ and\ \bibinfo {author} {\bibfnamefont {F.~F.}\ \bibnamefont
  {Assaad}},\ }\href {\doibase 10.1103/PhysRevB.87.075149} {\bibfield
  {journal} {\bibinfo  {journal} {Phys. Rev. B}\ }\textbf {\bibinfo {volume}
  {87}},\ \bibinfo {pages} {075149} (\bibinfo {year} {2013})}\BibitemShut
  {NoStop}%
\bibitem [{\citenamefont {Johnston}\ \emph {et~al.}(2013)\citenamefont
  {Johnston}, \citenamefont {Nowadnick}, \citenamefont {Kung}, \citenamefont
  {Moritz}, \citenamefont {Scalettar},\ and\ \citenamefont
  {Devereaux}}]{PhysRevB.87.235133}%
  \BibitemOpen
  \bibfield  {author} {\bibinfo {author} {\bibfnamefont {S.}~\bibnamefont
  {Johnston}}, \bibinfo {author} {\bibfnamefont {E.~A.}\ \bibnamefont
  {Nowadnick}}, \bibinfo {author} {\bibfnamefont {Y.~F.}\ \bibnamefont {Kung}},
  \bibinfo {author} {\bibfnamefont {B.}~\bibnamefont {Moritz}}, \bibinfo
  {author} {\bibfnamefont {R.~T.}\ \bibnamefont {Scalettar}}, \ and\ \bibinfo
  {author} {\bibfnamefont {T.~P.}\ \bibnamefont {Devereaux}},\ }\href {\doibase
  10.1103/PhysRevB.87.235133} {\bibfield  {journal} {\bibinfo  {journal} {Phys.
  Rev. B}\ }\textbf {\bibinfo {volume} {87}},\ \bibinfo {pages} {235133}
  (\bibinfo {year} {2013})}\BibitemShut {NoStop}%
\bibitem [{\citenamefont {Werner}\ and\ \citenamefont
  {Eckstein}(2013)}]{PhysRevB.88.165108}%
  \BibitemOpen
  \bibfield  {author} {\bibinfo {author} {\bibfnamefont {P.}~\bibnamefont
  {Werner}}\ and\ \bibinfo {author} {\bibfnamefont {M.}~\bibnamefont
  {Eckstein}},\ }\href {\doibase 10.1103/PhysRevB.88.165108} {\bibfield
  {journal} {\bibinfo  {journal} {Phys. Rev. B}\ }\textbf {\bibinfo {volume}
  {88}},\ \bibinfo {pages} {165108} (\bibinfo {year} {2013})}\BibitemShut
  {NoStop}%
\bibitem [{\citenamefont {Ohgoe}\ and\ \citenamefont
  {Imada}(2017)}]{PhysRevLett.119.197001}%
  \BibitemOpen
  \bibfield  {author} {\bibinfo {author} {\bibfnamefont {T.}~\bibnamefont
  {Ohgoe}}\ and\ \bibinfo {author} {\bibfnamefont {M.}~\bibnamefont {Imada}},\
  }\href {\doibase 10.1103/PhysRevLett.119.197001} {\bibfield  {journal}
  {\bibinfo  {journal} {Phys. Rev. Lett.}\ }\textbf {\bibinfo {volume} {119}},\
  \bibinfo {pages} {197001} (\bibinfo {year} {2017})}\BibitemShut {NoStop}%
\bibitem [{\citenamefont {Karakuzu}\ \emph {et~al.}(2017)\citenamefont
  {Karakuzu}, \citenamefont {Tocchio}, \citenamefont {Sorella},\ and\
  \citenamefont {Becca}}]{PhysRevB.96.205145}%
  \BibitemOpen
  \bibfield  {author} {\bibinfo {author} {\bibfnamefont {S.}~\bibnamefont
  {Karakuzu}}, \bibinfo {author} {\bibfnamefont {L.~F.}\ \bibnamefont
  {Tocchio}}, \bibinfo {author} {\bibfnamefont {S.}~\bibnamefont {Sorella}}, \
  and\ \bibinfo {author} {\bibfnamefont {F.}~\bibnamefont {Becca}},\ }\href
  {\doibase 10.1103/PhysRevB.96.205145} {\bibfield  {journal} {\bibinfo
  {journal} {Phys. Rev. B}\ }\textbf {\bibinfo {volume} {96}},\ \bibinfo
  {pages} {205145} (\bibinfo {year} {2017})}\BibitemShut {NoStop}%
\bibitem [{\citenamefont {Weber}\ and\ \citenamefont
  {Hohenadler}(2018)}]{PhysRevB.98.085405}%
  \BibitemOpen
  \bibfield  {author} {\bibinfo {author} {\bibfnamefont {M.}~\bibnamefont
  {Weber}}\ and\ \bibinfo {author} {\bibfnamefont {M.}~\bibnamefont
  {Hohenadler}},\ }\href {\doibase 10.1103/PhysRevB.98.085405} {\bibfield
  {journal} {\bibinfo  {journal} {Phys. Rev. B}\ }\textbf {\bibinfo {volume}
  {98}},\ \bibinfo {pages} {085405} (\bibinfo {year} {2018})}\BibitemShut
  {NoStop}%
\bibitem [{\citenamefont {Hohenadler}\ and\ \citenamefont
  {Fehske}(2018)}]{hohenadler2018density}%
  \BibitemOpen
  \bibfield  {author} {\bibinfo {author} {\bibfnamefont {M.}~\bibnamefont
  {Hohenadler}}\ and\ \bibinfo {author} {\bibfnamefont {H.}~\bibnamefont
  {Fehske}},\ }\href@noop {} {\bibfield  {journal} {\bibinfo  {journal} {The
  European Physical Journal B}\ }\textbf {\bibinfo {volume} {91}},\ \bibinfo
  {pages} {204} (\bibinfo {year} {2018})}\BibitemShut {NoStop}%
\bibitem [{\citenamefont {Karakuzu}\ \emph {et~al.}(2018)\citenamefont
  {Karakuzu}, \citenamefont {Seki},\ and\ \citenamefont
  {Sorella}}]{PhysRevB.98.201108}%
  \BibitemOpen
  \bibfield  {author} {\bibinfo {author} {\bibfnamefont {S.}~\bibnamefont
  {Karakuzu}}, \bibinfo {author} {\bibfnamefont {K.}~\bibnamefont {Seki}}, \
  and\ \bibinfo {author} {\bibfnamefont {S.}~\bibnamefont {Sorella}},\ }\href
  {\doibase 10.1103/PhysRevB.98.201108} {\bibfield  {journal} {\bibinfo
  {journal} {Phys. Rev. B}\ }\textbf {\bibinfo {volume} {98}},\ \bibinfo
  {pages} {201108} (\bibinfo {year} {2018})}\BibitemShut {NoStop}%
\bibitem [{\citenamefont {Wang}\ \emph {et~al.}(2019)\citenamefont {Wang},
  \citenamefont {Esterlis}, \citenamefont {Shi}, \citenamefont {Cirac},\ and\
  \citenamefont {Demler}}]{wang2019zero}%
  \BibitemOpen
  \bibfield  {author} {\bibinfo {author} {\bibfnamefont {Y.}~\bibnamefont
  {Wang}}, \bibinfo {author} {\bibfnamefont {I.}~\bibnamefont {Esterlis}},
  \bibinfo {author} {\bibfnamefont {T.}~\bibnamefont {Shi}}, \bibinfo {author}
  {\bibfnamefont {J.~I.}\ \bibnamefont {Cirac}}, \ and\ \bibinfo {author}
  {\bibfnamefont {E.}~\bibnamefont {Demler}},\ }\href@noop {} {\bibfield
  {journal} {\bibinfo  {journal} {arXiv preprint arXiv:1910.01792}\ } (\bibinfo
  {year} {2019})}\BibitemShut {NoStop}%
\bibitem [{\citenamefont {Kazuhiro}\ \emph {et~al.}(2020)\citenamefont
  {Kazuhiro}, \citenamefont {Seiji}, \citenamefont {Sandro} \emph
  {et~al.}}]{kazuhiro2020phase}%
  \BibitemOpen
  \bibfield  {author} {\bibinfo {author} {\bibfnamefont {S.}~\bibnamefont
  {Kazuhiro}}, \bibinfo {author} {\bibfnamefont {Y.}~\bibnamefont {Seiji}},
  \bibinfo {author} {\bibfnamefont {S.}~\bibnamefont {Sandro}},  \emph
  {et~al.},\ }\href@noop {} {\bibfield  {journal} {\bibinfo  {journal}
  {Communications Physics}\ }\textbf {\bibinfo {volume} {3}} (\bibinfo {year}
  {2020})}\BibitemShut {NoStop}%
\bibitem [{\citenamefont {Li}\ and\ \citenamefont {Yao}(2019)}]{Yao-review}%
  \BibitemOpen
  \bibfield  {author} {\bibinfo {author} {\bibfnamefont {Z.-X.}\ \bibnamefont
  {Li}}\ and\ \bibinfo {author} {\bibfnamefont {H.}~\bibnamefont {Yao}},\
  }\href {\doibase 10.1146/annurev-conmatphys-033117-054307} {\bibfield
  {journal} {\bibinfo  {journal} {Annual Review of Condensed Matter Physics}\
  }\textbf {\bibinfo {volume} {10}},\ \bibinfo {pages} {337} (\bibinfo {year}
  {2019})}\BibitemShut {NoStop}%
\bibitem [{\citenamefont {Hohenadler}\ and\ \citenamefont {von~der
  Linden}(2007)}]{Hohenadler2007}%
  \BibitemOpen
  \bibfield  {author} {\bibinfo {author} {\bibfnamefont {M.}~\bibnamefont
  {Hohenadler}}\ and\ \bibinfo {author} {\bibfnamefont {W.}~\bibnamefont
  {von~der Linden}},\ }\enquote {\bibinfo {title} {Lang-firsov approaches to
  polaron physics: From variational methods to unbiased quantum monte carlo
  simulations},}\ in\ \href {\doibase 10.1007/978-1-4020-6348-0_11} {\emph
  {\bibinfo {booktitle} {Polarons in Advanced Materials}}},\ \bibinfo {editor}
  {edited by\ \bibinfo {editor} {\bibfnamefont {A.~S.}\ \bibnamefont
  {Alexandrov}}}\ (\bibinfo  {publisher} {Springer Netherlands},\ \bibinfo
  {address} {Dordrecht},\ \bibinfo {year} {2007})\ pp.\ \bibinfo {pages}
  {463--502}\BibitemShut {NoStop}%
\bibitem [{\citenamefont {Lax}(1952)}]{lax1952franck}%
  \BibitemOpen
  \bibfield  {author} {\bibinfo {author} {\bibfnamefont {M.}~\bibnamefont
  {Lax}},\ }\href@noop {} {\bibfield  {journal} {\bibinfo  {journal} {The
  Journal of chemical physics}\ }\textbf {\bibinfo {volume} {20}},\ \bibinfo
  {pages} {1752} (\bibinfo {year} {1952})}\BibitemShut {NoStop}%
\bibitem [{\citenamefont {Kivelson}\ \emph {et~al.}(1990)\citenamefont
  {Kivelson}, \citenamefont {Emery},\ and\ \citenamefont
  {Lin}}]{PhysRevB.42.6523}%
  \BibitemOpen
  \bibfield  {author} {\bibinfo {author} {\bibfnamefont {S.~A.}\ \bibnamefont
  {Kivelson}}, \bibinfo {author} {\bibfnamefont {V.~J.}\ \bibnamefont {Emery}},
  \ and\ \bibinfo {author} {\bibfnamefont {H.~Q.}\ \bibnamefont {Lin}},\ }\href
  {\doibase 10.1103/PhysRevB.42.6523} {\bibfield  {journal} {\bibinfo
  {journal} {Phys. Rev. B}\ }\textbf {\bibinfo {volume} {42}},\ \bibinfo
  {pages} {6523} (\bibinfo {year} {1990})}\BibitemShut {NoStop}%
\bibitem [{\citenamefont {Green}\ \emph {et~al.}(2018)\citenamefont {Green},
  \citenamefont {Conduit},\ and\ \citenamefont
  {Krüger}}]{doi:10.1146/annurev-conmatphys-033117-053925}%
  \BibitemOpen
  \bibfield  {author} {\bibinfo {author} {\bibfnamefont {A.~G.}\ \bibnamefont
  {Green}}, \bibinfo {author} {\bibfnamefont {G.}~\bibnamefont {Conduit}}, \
  and\ \bibinfo {author} {\bibfnamefont {F.}~\bibnamefont {Krüger}},\ }\href
  {\doibase 10.1146/annurev-conmatphys-033117-053925} {\bibfield  {journal}
  {\bibinfo  {journal} {Annual Review of Condensed Matter Physics}\ }\textbf
  {\bibinfo {volume} {9}},\ \bibinfo {pages} {59} (\bibinfo {year} {2018})},\
  \Eprint
  {http://arxiv.org/abs/https://doi.org/10.1146/annurev-conmatphys-033117-053925}
  {https://doi.org/10.1146/annurev-conmatphys-033117-053925} \BibitemShut
  {NoStop}%
\bibitem [{\citenamefont {Kohn}\ and\ \citenamefont
  {Luttinger}(1965)}]{PhysRevLett.15.524}%
  \BibitemOpen
  \bibfield  {author} {\bibinfo {author} {\bibfnamefont {W.}~\bibnamefont
  {Kohn}}\ and\ \bibinfo {author} {\bibfnamefont {J.~M.}\ \bibnamefont
  {Luttinger}},\ }\href {\doibase 10.1103/PhysRevLett.15.524} {\bibfield
  {journal} {\bibinfo  {journal} {Phys. Rev. Lett.}\ }\textbf {\bibinfo
  {volume} {15}},\ \bibinfo {pages} {524} (\bibinfo {year} {1965})}\BibitemShut
  {NoStop}%
\bibitem [{\citenamefont {MacCluer}(2000)}]{maccluer2000many}%
  \BibitemOpen
  \bibfield  {author} {\bibinfo {author} {\bibfnamefont {C.~R.}\ \bibnamefont
  {MacCluer}},\ }\href@noop {} {\bibfield  {journal} {\bibinfo  {journal} {Siam
  Review}\ }\textbf {\bibinfo {volume} {42}},\ \bibinfo {pages} {487} (\bibinfo
  {year} {2000})}\BibitemShut {NoStop}%
\bibitem [{\citenamefont {Agterberg}\ \emph {et~al.}(2020)\citenamefont
  {Agterberg}, \citenamefont {Davis}, \citenamefont {Edkins}, \citenamefont
  {Fradkin}, \citenamefont {Van~Harlingen}, \citenamefont {Kivelson},
  \citenamefont {Lee}, \citenamefont {Radzihovsky}, \citenamefont {Tranquada},\
  and\ \citenamefont {Wang}}]{agterberg2020physics}%
  \BibitemOpen
  \bibfield  {author} {\bibinfo {author} {\bibfnamefont {D.~F.}\ \bibnamefont
  {Agterberg}}, \bibinfo {author} {\bibfnamefont {J.~S.}\ \bibnamefont
  {Davis}}, \bibinfo {author} {\bibfnamefont {S.~D.}\ \bibnamefont {Edkins}},
  \bibinfo {author} {\bibfnamefont {E.}~\bibnamefont {Fradkin}}, \bibinfo
  {author} {\bibfnamefont {D.~J.}\ \bibnamefont {Van~Harlingen}}, \bibinfo
  {author} {\bibfnamefont {S.~A.}\ \bibnamefont {Kivelson}}, \bibinfo {author}
  {\bibfnamefont {P.~A.}\ \bibnamefont {Lee}}, \bibinfo {author} {\bibfnamefont
  {L.}~\bibnamefont {Radzihovsky}}, \bibinfo {author} {\bibfnamefont {J.~M.}\
  \bibnamefont {Tranquada}}, \ and\ \bibinfo {author} {\bibfnamefont
  {Y.}~\bibnamefont {Wang}},\ }\href@noop {} {\bibfield  {journal} {\bibinfo
  {journal} {Annual Review of Condensed Matter Physics}\ }\textbf {\bibinfo
  {volume} {11}},\ \bibinfo {pages} {231} (\bibinfo {year} {2020})}\BibitemShut
  {NoStop}%
\bibitem [{\citenamefont {You}\ \emph {et~al.}(2012)\citenamefont {You},
  \citenamefont {Chen}, \citenamefont {Sun},\ and\ \citenamefont
  {Zhai}}]{HuiZhai_2012}%
  \BibitemOpen
  \bibfield  {author} {\bibinfo {author} {\bibfnamefont {Y.-Z.}\ \bibnamefont
  {You}}, \bibinfo {author} {\bibfnamefont {Z.}~\bibnamefont {Chen}}, \bibinfo
  {author} {\bibfnamefont {X.-Q.}\ \bibnamefont {Sun}}, \ and\ \bibinfo
  {author} {\bibfnamefont {H.}~\bibnamefont {Zhai}},\ }\href {\doibase
  10.1103/PhysRevLett.109.265302} {\bibfield  {journal} {\bibinfo  {journal}
  {Phys. Rev. Lett.}\ }\textbf {\bibinfo {volume} {109}},\ \bibinfo {pages}
  {265302} (\bibinfo {year} {2012})}\BibitemShut {NoStop}%
\bibitem [{\citenamefont {Schmid}\ \emph {et~al.}(2002)\citenamefont {Schmid},
  \citenamefont {Todo}, \citenamefont {Troyer},\ and\ \citenamefont
  {Dorneich}}]{PhysRevLett.88.167208}%
  \BibitemOpen
  \bibfield  {author} {\bibinfo {author} {\bibfnamefont {G.}~\bibnamefont
  {Schmid}}, \bibinfo {author} {\bibfnamefont {S.}~\bibnamefont {Todo}},
  \bibinfo {author} {\bibfnamefont {M.}~\bibnamefont {Troyer}}, \ and\ \bibinfo
  {author} {\bibfnamefont {A.}~\bibnamefont {Dorneich}},\ }\href {\doibase
  10.1103/PhysRevLett.88.167208} {\bibfield  {journal} {\bibinfo  {journal}
  {Phys. Rev. Lett.}\ }\textbf {\bibinfo {volume} {88}},\ \bibinfo {pages}
  {167208} (\bibinfo {year} {2002})}\BibitemShut {NoStop}%
\bibitem [{\citenamefont {Chen}\ \emph {et~al.}(2008)\citenamefont {Chen},
  \citenamefont {Melko}, \citenamefont {Wessel},\ and\ \citenamefont
  {Kao}}]{PhysRevB.77.014524}%
  \BibitemOpen
  \bibfield  {author} {\bibinfo {author} {\bibfnamefont {Y.-C.}\ \bibnamefont
  {Chen}}, \bibinfo {author} {\bibfnamefont {R.~G.}\ \bibnamefont {Melko}},
  \bibinfo {author} {\bibfnamefont {S.}~\bibnamefont {Wessel}}, \ and\ \bibinfo
  {author} {\bibfnamefont {Y.-J.}\ \bibnamefont {Kao}},\ }\href {\doibase
  10.1103/PhysRevB.77.014524} {\bibfield  {journal} {\bibinfo  {journal} {Phys.
  Rev. B}\ }\textbf {\bibinfo {volume} {77}},\ \bibinfo {pages} {014524}
  (\bibinfo {year} {2008})}\BibitemShut {NoStop}%
\bibitem [{\citenamefont {Melko}\ \emph {et~al.}(2005)\citenamefont {Melko},
  \citenamefont {Paramekanti}, \citenamefont {Burkov}, \citenamefont
  {Vishwanath}, \citenamefont {Sheng},\ and\ \citenamefont
  {Balents}}]{PhysRevLett.95.127207}%
  \BibitemOpen
  \bibfield  {author} {\bibinfo {author} {\bibfnamefont {R.~G.}\ \bibnamefont
  {Melko}}, \bibinfo {author} {\bibfnamefont {A.}~\bibnamefont {Paramekanti}},
  \bibinfo {author} {\bibfnamefont {A.~A.}\ \bibnamefont {Burkov}}, \bibinfo
  {author} {\bibfnamefont {A.}~\bibnamefont {Vishwanath}}, \bibinfo {author}
  {\bibfnamefont {D.~N.}\ \bibnamefont {Sheng}}, \ and\ \bibinfo {author}
  {\bibfnamefont {L.}~\bibnamefont {Balents}},\ }\href {\doibase
  10.1103/PhysRevLett.95.127207} {\bibfield  {journal} {\bibinfo  {journal}
  {Phys. Rev. Lett.}\ }\textbf {\bibinfo {volume} {95}},\ \bibinfo {pages}
  {127207} (\bibinfo {year} {2005})}\BibitemShut {NoStop}%
\bibitem [{\citenamefont {Wang}\ \emph {et~al.}(2009)\citenamefont {Wang},
  \citenamefont {Pollmann},\ and\ \citenamefont
  {Vishwanath}}]{PhysRevLett.102.017203}%
  \BibitemOpen
  \bibfield  {author} {\bibinfo {author} {\bibfnamefont {F.}~\bibnamefont
  {Wang}}, \bibinfo {author} {\bibfnamefont {F.}~\bibnamefont {Pollmann}}, \
  and\ \bibinfo {author} {\bibfnamefont {A.}~\bibnamefont {Vishwanath}},\
  }\href {\doibase 10.1103/PhysRevLett.102.017203} {\bibfield  {journal}
  {\bibinfo  {journal} {Phys. Rev. Lett.}\ }\textbf {\bibinfo {volume} {102}},\
  \bibinfo {pages} {017203} (\bibinfo {year} {2009})}\BibitemShut {NoStop}%
\bibitem [{\citenamefont {Wessel}(2007)}]{PhysRevB.75.174301}%
  \BibitemOpen
  \bibfield  {author} {\bibinfo {author} {\bibfnamefont {S.}~\bibnamefont
  {Wessel}},\ }\href {\doibase 10.1103/PhysRevB.75.174301} {\bibfield
  {journal} {\bibinfo  {journal} {Phys. Rev. B}\ }\textbf {\bibinfo {volume}
  {75}},\ \bibinfo {pages} {174301} (\bibinfo {year} {2007})}\BibitemShut
  {NoStop}%
\bibitem [{\citenamefont {Li}\ \emph {et~al.}(2019)\citenamefont {Li},
  \citenamefont {Cohen},\ and\ \citenamefont {Lee}}]{PhysRevB.100.245105}%
  \BibitemOpen
  \bibfield  {author} {\bibinfo {author} {\bibfnamefont {Z.-X.}\ \bibnamefont
  {Li}}, \bibinfo {author} {\bibfnamefont {M.~L.}\ \bibnamefont {Cohen}}, \
  and\ \bibinfo {author} {\bibfnamefont {D.-H.}\ \bibnamefont {Lee}},\ }\href
  {\doibase 10.1103/PhysRevB.100.245105} {\bibfield  {journal} {\bibinfo
  {journal} {Phys. Rev. B}\ }\textbf {\bibinfo {volume} {100}},\ \bibinfo
  {pages} {245105} (\bibinfo {year} {2019})}\BibitemShut {NoStop}%
\bibitem [{\citenamefont {MacDonald}\ \emph {et~al.}(1988)\citenamefont
  {MacDonald}, \citenamefont {Girvin},\ and\ \citenamefont
  {Yoshioka}}]{PhysRevB.37.9753}%
  \BibitemOpen
  \bibfield  {author} {\bibinfo {author} {\bibfnamefont {A.~H.}\ \bibnamefont
  {MacDonald}}, \bibinfo {author} {\bibfnamefont {S.~M.}\ \bibnamefont
  {Girvin}}, \ and\ \bibinfo {author} {\bibfnamefont {D.}~\bibnamefont
  {Yoshioka}},\ }\href {\doibase 10.1103/PhysRevB.37.9753} {\bibfield
  {journal} {\bibinfo  {journal} {Phys. Rev. B}\ }\textbf {\bibinfo {volume}
  {37}},\ \bibinfo {pages} {9753} (\bibinfo {year} {1988})}\BibitemShut
  {NoStop}%
\bibitem [{\citenamefont {de~Oliveira}(1993)}]{PhysRevB.48.6141}%
  \BibitemOpen
  \bibfield  {author} {\bibinfo {author} {\bibfnamefont {M.~J.}\ \bibnamefont
  {de~Oliveira}},\ }\href {\doibase 10.1103/PhysRevB.48.6141} {\bibfield
  {journal} {\bibinfo  {journal} {Phys. Rev. B}\ }\textbf {\bibinfo {volume}
  {48}},\ \bibinfo {pages} {6141} (\bibinfo {year} {1993})}\BibitemShut
  {NoStop}%
\bibitem [{\citenamefont {Capriotti}\ \emph {et~al.}(1999)\citenamefont
  {Capriotti}, \citenamefont {Trumper},\ and\ \citenamefont
  {Sorella}}]{PhysRevLett.82.3899}%
  \BibitemOpen
  \bibfield  {author} {\bibinfo {author} {\bibfnamefont {L.}~\bibnamefont
  {Capriotti}}, \bibinfo {author} {\bibfnamefont {A.~E.}\ \bibnamefont
  {Trumper}}, \ and\ \bibinfo {author} {\bibfnamefont {S.}~\bibnamefont
  {Sorella}},\ }\href {\doibase 10.1103/PhysRevLett.82.3899} {\bibfield
  {journal} {\bibinfo  {journal} {Phys. Rev. Lett.}\ }\textbf {\bibinfo
  {volume} {82}},\ \bibinfo {pages} {3899} (\bibinfo {year}
  {1999})}\BibitemShut {NoStop}%
\bibitem [{\citenamefont {Reger}\ \emph {et~al.}(1989)\citenamefont {Reger},
  \citenamefont {Riera},\ and\ \citenamefont {Young}}]{Reger_1989}%
  \BibitemOpen
  \bibfield  {author} {\bibinfo {author} {\bibfnamefont {J.~D.}\ \bibnamefont
  {Reger}}, \bibinfo {author} {\bibfnamefont {J.~A.}\ \bibnamefont {Riera}}, \
  and\ \bibinfo {author} {\bibfnamefont {A.~P.}\ \bibnamefont {Young}},\ }\href
  {\doibase 10.1088/0953-8984/1/10/007} {\bibfield  {journal} {\bibinfo
  {journal} {Journal of Physics: Condensed Matter}\ }\textbf {\bibinfo {volume}
  {1}},\ \bibinfo {pages} {1855} (\bibinfo {year} {1989})}\BibitemShut
  {NoStop}%
\bibitem [{\citenamefont {Evenbly}\ and\ \citenamefont
  {Vidal}(2010)}]{PhysRevLett.104.187203}%
  \BibitemOpen
  \bibfield  {author} {\bibinfo {author} {\bibfnamefont {G.}~\bibnamefont
  {Evenbly}}\ and\ \bibinfo {author} {\bibfnamefont {G.}~\bibnamefont
  {Vidal}},\ }\href {\doibase 10.1103/PhysRevLett.104.187203} {\bibfield
  {journal} {\bibinfo  {journal} {Phys. Rev. Lett.}\ }\textbf {\bibinfo
  {volume} {104}},\ \bibinfo {pages} {187203} (\bibinfo {year}
  {2010})}\BibitemShut {NoStop}%
\end{thebibliography}%

\newpage
\onecolumngrid

\section*{Supplemental Material}
\subsection{A. Momentum Space Path Integral Derivation of generalized Holstein-Lang-Firsov transformation}
  We consider the electron-phonon problem in a general scenario, where we allow arbitrary phonon dispersion and electronic band structure and interaction, with the only assumption that e-ph interaction couples phonon coordinates and electron densities. Writing the phonon degrees of freedom in their normal modes, we have Hamiltonian:
\begin{align}
\hat{H}_{\text{e}} &= - \sum_{ij,\sigma}  (t_{ij}\hat{c
}_{i,\sigma}^{\dagger} \hat{c}_{j,\sigma} + \text{h.c.}) + \frac{1}{2}\sum_{ij} U^{\text{e-e}}_{ij} \hat{n}_i\hat{n}_j \\
\hat{H}_{\text{ph}} &=  \sum_{l}\left( \frac{k_l \hat{x}_l^2 }{2} + \frac{\hat{p}_l^2}{2m_l} \right) \\
\hat{H}_{\text{e-ph}} &= \sum_{i,l}\alpha_{il} \hat{n}_i \hat{x}_l
\end{align}

This problem have an alternative form by a generalized Hostein-Lang-Firsov transformation. To see this, we perform path integral tracking the phonon degrees of freedom in their momentum space. After Trotter-Suzuki decomposition, the phonon-related terms at imaginary time $\tau$ can be evaluated:
\begin{align}
& \int \prod_l \mathrm{d}p^\tau_l \mathrm{d}x^\tau_l \langle p^\tau_l  |\mathrm{e}^{-\Delta \tau (\frac{\hat{p}_l^2}{2m}+\frac{k\hat{x}^2_l}{2}+\sum_i \alpha_{il} \hat{n}_i \hat{x}_l)}|x^\tau_l \rangle\langle x^\tau_l|p^{\tau+\Delta\tau}_{l+1}\rangle \nonumber\\
=& \int \prod_l  \mathrm{d}p^\tau_l \mathrm{d}x_l \mathrm{e}^{-\Delta \tau \left[\frac{p_l^2}{2m}+\frac{k_l x^2_l}{2}+(\sum_i \alpha_{il} \hat{n}_l - \mathrm{i}\dot{p}_l)x_l\right]} \nonumber\\
\propto & \int \prod_l  \mathrm{d}p^\tau_l  \exp \left\{-\Delta \tau \left[\sum_{l} (\frac{p_l^2}{2m}+\frac{\dot{p}_l^2}{2k_l}) - \sum_{ij} \frac{U^{\text{e-ph}}_{ij}}{2} \hat{n}_i\hat{n}_j + \sum_{i,l} \mathrm{i}\frac{\alpha_{il}}{k_l} \hat{n}_i \dot{p}_l \right]\right\}
\end{align}
where $U^{\text{e-ph}}_{ij} \equiv \sum_l \alpha_{il}\alpha_{jl}/k_l$. After performing the fermionic coherent state path integral, we reach the action:
\begin{align}
Z & = \int \mathcal{D}[\Bar{\psi}_i, \psi_i]\mathcal{D}[p_i] \mathrm{e}^{-S[\Bar{\psi}_i, \psi_i;p_i]}\nonumber\\
S & = \int \mathrm{d}\tau \sum_{ij} \Bar{\psi}_{i\sigma}\left[(\partial_\tau -\mu + \sum_l\mathrm{i}\frac{\alpha_{il}}{k_l}\dot{p}_l)\delta_{ij} - t_{ij}\right] \psi_{j\sigma}+ \sum_{ij} \frac{U^{\text{eff}}_{ij}}{2}n_{i} n_j + \sum_l (\frac{p_l^2}{2m} + \frac{\dot{p}_i^2}{2k_l})\nonumber\\
& = \int \mathrm{d}\tau \sum_{ij} \Bar{\psi}_{i\sigma}\left[(\partial_\tau -\mu )\delta_{ij} - \Tilde{t}_{ij}\right] \psi_{j\sigma}+\sum_{ij} \frac{U^{\text{eff}}_{ij}}{2}n_{i} n_j + \sum_l (\frac{p_l^2}{2m} + \frac{\dot{p}_i^2}{2k_l}) 
\end{align}
where $ \Tilde{t}_{ij} \equiv t_{ij} \mathrm{e}^{ \sum_l \mathrm{i}\frac{\alpha_{il}-\alpha_{jl}}{k_l}p_l}$ and $U^{\text{eff}}_{ij} = U^{\text{e-e}}_{ij}-U^{\text{e-ph}}_{ij}$. In the last line we perform unitary transformation $\psi_{i\sigma} \to \psi_{i\sigma} \mathrm{e}^{-\sum_l \mathrm{i} \frac{\alpha_{il}}{k_l}p_l}$. (For $|U^{\text{eff}}_{ij}|<|t_{ij}|$, this form of action may be useful for perturbation theory.) This expression is exactly the momentum space path integral of the transformed theory by the generalized unitary transformation $
\hat{U} = \exp\left[\mathrm{i}\sum_{j,l}\alpha_{jl} k_l^{-1} \hat{n}_j \hat{p}_l\right]$:
\begin{align}
\hat{H} &= - \sum_{ij,\sigma}  (\Tilde{t}_{ij}\hat{c
}_{i,\sigma}^{\dagger} \hat{c}_{j,\sigma} + \text{h.c.}) + \frac{1}{2}\sum_{ij} U^{\text{eff}}_{ij} \hat{n}_i\hat{n}_j+ \sum_{l}\left( \frac{k_l \hat{x}_l^2 }{2} + \frac{\hat{p}_l^2}{2m_l} \right)
\end{align}

For sound modes, translation symmetry of the whole crystal makes the coupling strength between electron density and a sound mode $\Vec{u}$ on wavevector $\Vec{q}$ at most $\alpha(\Vec{q})\sim \Vec{q}\cdot\Vec{u}$ and the normal mode stiffness $k(\Vec{q}) \sim q^2$ in the infrared limit, which results in a quasi-long-range $\sim r^{-d}$ interaction at long distance in $d$-dimension. While optical modes with $k(\Vec{q}) \sim q_0^2 + q^2$ and $\alpha(\Vec{q}) \sim \alpha_0 + \alpha' q^{\gamma \geq0}$ generate exponentially decaying interaction. 

After taking dispersion of phonons and more generic coupling into consideration, we will have a complicated Hamiltonian at the zeroth order of expansion at strong-coupling limit. Truncating the interactions that is weaker than the leading quantum hopping, we can first solve a classical interacting lattice electron gas problem and then introduce hopping and longer range interactions as perturbations. Therefore, in seeking a charge $(n>2)e$ superconductor, we may resort to a weakly dispersive or widerly coupled optical mode that generate relatively weak nearest-neighbor ($\sim \frac{t^2}{U_{\text{e-e}}}$) attraction to tune the effective $V/J$ into the desirable range. As a concrete example, adding weak phonon coupling to nearest neighbor sites in the original Holstein-Hubbard model:
\begin{align}
\hat{H}' &= \sum_{<i,j>} \alpha' (\hat{n}_i + \hat{n}_j) (\hat{x}_i + \hat{x}_j)
\end{align}
can introduce small on-site and nearest-neighbor attraction $\sim \frac{\alpha'^2}{k}$ without modifying the hopping term.

\subsection{B. Sign-problem free on bipartite lattices at half filling and $U_{\text{eff}}\geq 0$}
At half filling, $\mu =0 $ and the interacting term in the transformed Hamiltonian can be written as
\begin{align} \label{half}
U_{\text{eff}} \sum_i (\hat{n}_{i\uparrow}-\frac{1}{2})(\hat{n}_{i\downarrow}-\frac{1}{2})
\end{align}
Applying discrete Hubbard-Stratonovich transformation, and defining $\lambda = \cosh^{-1}(\mathrm{e}^{\Delta \tau U_{\text{eff}}/2})$:
\begin{align}
\mathrm{e}^{-\Delta\tau U_{
\text{eff}} (\hat{n}_{i\uparrow}-\frac{1}{2})(\hat{n}_{i\downarrow}-\frac{1}{2})} = \frac{1}{2}\mathrm{e}^{-\Delta\tau U_{
\text{eff}}/4} \sum_{s=\pm 1} \mathrm{e}^{-\lambda s (\hat{n}_{i\uparrow} - \hat{n}_{i\downarrow})}
\end{align}

Under $\hat{c}^{\dagger}_{i\downarrow} \rightarrow (-1)^i  \hat{c}_{i\downarrow}$ on bipartite lattice:
\begin{align} \label{trans}
\sum_{\sigma} (\hat{S}_{ij}\hat{c
}_{i,\sigma}^{\dagger} \hat{c}_{j,\sigma} + \text{h.c.}) &\rightarrow (\hat{S}_{ij}\hat{c
}_{i,\uparrow}^{\dagger} \hat{c}_{j,\uparrow}+ \hat{S}^{\dagger}_{ij}\hat{c
}_{i,\downarrow}^{\dagger} \hat{c}_{j,\downarrow} + \text{h.c.}) \nonumber\\
\mathrm{e}^{-\lambda s (\hat{n}_{i\uparrow} - \hat{n}_{i\downarrow})} &\rightarrow \mathrm{e}^{-\lambda s (\hat{n}_{i\uparrow} + \hat{n}_{i\downarrow})}
\end{align}

In momentum basis, $S_{ij}= \mathrm{e}^{\mathrm{i}\frac{\alpha}{k}(p_j-p_i)}$ is a pure phase for arbitrary phonon configurations. Therefore the determinants for up- and down-spin are mutually complex conjugated; the problem is thus sign free. Another approach rendering this model sign-problem free appears in Ref.~\cite{PhysRevB.98.201108}.

\subsection{C. Detailed Calculation of $F$, $F'$}

Take the computation for $J = \frac{t^2}{U_\text{eff}}F(X,Y)$  as an example ($t_b$, $V_b$ can be similarly acquired), we consider two singly occupied sites and the virtual process in which  electron hopping between the two sites occurs twice. In the form factor all possible phonon configurations on two sites in the intermediate state should be considered:
\begin{align}
F(X,Y) &= \frac{U_\text{eff}}{\omega_D}\sum_{n,m=0}^{\infty} \frac{\langle 0,0| \exp{\left[-\mathrm{i}(\alpha/k)(\hat{p}_1 - \hat{p}_2)\right]}|n,m\rangle\langle n,m| \exp{\left[\mathrm{i}(\alpha/k)(\hat{p}_1 - \hat{p}_2)\right]}|0,0\rangle}{(n+m)+\frac{U_\text{eff}}{\omega_D}}\nonumber\\
&= Y \sum_{n,m=0}^{\infty} \frac{\langle 0| \exp{\left[-\sqrt{\frac{X}{2}}(a - a^{\dagger})\right]}|n\rangle\langle n|\exp{\left[\sqrt{\frac{X}{2}}(a - a^{\dagger})\right]}|0\rangle \cdot (n\leftrightarrow m, \text{h.c.})}{(n+m)+Y}\nonumber\\
&= Y \int_0^{\infty} \mathrm{d} t  \sum_{n,m=0}^{\infty} \exp{\left[-(n+m)t-Y t\right]} \nonumber\\
& \ \ \ \ \ \ \ \cdot\langle 0| \exp{\left[-\sqrt{\frac{X}{2}}(a - a^{\dagger})\right]}|n\rangle\langle n|\exp{\left[\sqrt{\frac{X}{2}}(a - a^{\dagger})\right]}|0\rangle \cdot (n\leftrightarrow m, \text{h.c.})\nonumber\\
&= Y \int_0^{\infty} \mathrm{d} t  \sum_{n,m=0}^{\infty}  \exp{\left[-(n+m)t- Y t-X\right]}\cdot\frac{(\frac{X}{2})^n}{n!}\cdot\frac{(\frac{X}{2})^m}{m!}\nonumber\\
&=Y \mathrm{e}^{-X} \int_0^{\infty} \mathrm{d} t \exp{\left\{-Y t+ X\exp(-t)\right\}} \nonumber\\
&= Y \mathrm{e}^{-X} \int_0^1 \mathrm{d}z \cdot z^{Y-1} \mathrm{e}^{X z}
\end{align}
The last line is easier for numerical integration. 

Similarly we can compute $F'$, with the difference that $U_{\text{eff}}$ doesn't play a role and the intermediate state must have phonon. Taking $V=\frac{t^2}{U_\text{e-ph}}F'(X)$ as an example:
\begin{align}
F'(X) &= X \sum_{n+m\neq0}^{\infty} \frac{\langle 0| \exp{\left[-\sqrt{\frac{X}{2}}(a - a^{\dagger})\right]}|n\rangle\langle n|\exp{\left[\sqrt{\frac{X}{2}}(a - a^{\dagger})\right]}|0\rangle \cdot (n\leftrightarrow m, \text{h.c.})}{n+m}\nonumber\\
&= X\int_0^{\infty} \mathrm{d} t  \sum_{n+m\neq0}^{\infty} \exp{\left[-(n+m)t-X\right]} \cdot\frac{(\frac{X}{2})^n}{n!}\cdot\frac{(\frac{X}{2})^m}{m!}\nonumber\\
&= X\mathrm{e}^{-X} \int_0^{\infty} \mathrm{d} t  \cdot(\mathrm{e}^{ X\mathrm{e}^{-t}}-1) \nonumber\\
&=X\mathrm{e}^{-X} \int_0^{1} \mathrm{d} z  \cdot z^{-1}(\mathrm{e}^{ X z}-1)
\end{align}
Alternatively, the primed factors can also be computed via identity $F' = X \frac{\partial F}{\partial Y}|_{Y\rightarrow 0}$. 

The calculation for $t_2$ and $\tau$ are analogous, except that the intermediate state could excite phonon only on the middle site.

It is useful to consider an approximate expression of $F(x,y)$ to investigate the crossover between two limiting behaviors (assume $x>0$):
\begin{align}
F(x,y) &= \int_0^{\infty} \mathrm{d} t \exp{\left\{-t-x\left[1- \exp(-t/y)\right]\right\}} \nonumber\\
&\geq \int_0^{y} \mathrm{d} t \exp{\left\{-t-\frac{x}{y} t\right\}} +\int_y^\infty \mathrm{d} t \exp{\left\{-t-x\right\}} \nonumber\\
&= \frac{y}{x+y} +\frac{x}{x+y}\mathrm{e}^{-(x+y)} \label{approx} \\
F(-x,y)
&\leq \frac{x}{x-y}\mathrm{e}^{-(x+y)} - \frac{y}{x-y} \mathrm{e}^{-2x}
\end{align}
In practice these expressions give quite good approximation to the original integral and is exact in adiabatic and anti-adiabatic limit. 

From this expression we can estimate that, when $X\mathrm{e}^{-(X+Y)} \ll Y$, or equivalently $X\gg \frac{U_{\text{e-ph}}}{U_{\text{e-e}}} \ln \frac{U_{\text{e-ph}}}{U_{\text{eff}}}$, $F(X,Y)$ crossover from $1$ to  $\frac{Y}{X+Y}$. When $U_{\text{eff}}/U_{\text{e-e}}$ is small, this crossover boundary approximates to $U_{\text{eff}} \sim U_{\text{e-e}} \mathrm{e}^{-X}$.

\subsection{D. Finite-temperature correction to $F$'s}

To take temperature into account, we replace all $\langle0|\cdot|0\rangle$ with $(1-\mathrm{e}^{-\beta \omega_D})\sum_k\mathrm{e}^{-k\beta \omega_D} \langle k|\cdot|k\rangle $. Then we will need to evaluate terms like ($D$ is the displacement operator of boson coherent states):
\begin{align}
\sum_{k,n}\langle n| D(\pm a) |k\rangle \mathrm{e}^{-k\epsilon_1} \langle k| D(a) |n\rangle \mathrm{e}^{-n\epsilon_2}=\exp \left[-a^2\frac{(1\pm\mathrm{e}^{-\epsilon_1})(1\pm\mathrm{e}^{-\epsilon_1})}{1-\mathrm{e}^{-\epsilon_1}\mathrm{e}^{-\epsilon_2}}\right]
\end{align}
With the aid of this identity and $\frac{1}{n+Y} = \frac{Y}{2\sin \pi Y} \int_{-\pi}^{\pi} \mathrm{d}\theta (-)^n \mathrm{e}^{\mathrm{i}(n+Y)\theta}$ , we might find the modified $F$'s (we label $F(X,Y)$ as $F_{\text{sgn}(X)}(|X|,Y)$ to account for the different behavior when $X$ take different signs):
\begin{align}
F_{\beta,\pm(X,Y)} &= \frac{Y}{2\sin(\pi Y)}\int_{-\pi}^{\pi} \mathrm{d} \theta \exp{\left\{\mathrm{i}Y\theta-X\left[
\frac{(1\pm \mathrm{e}^{-\beta\omega_D-\mathrm{i}\theta})(1\pm \mathrm{e}^{\mathrm{i}\theta})}{1-\mathrm{e}^{-\beta\omega_D}}\right]\right\}} \nonumber\\
&=\frac{Y}{\sin(\pi Y)}\int_{0}^{\pi} \mathrm{d} \theta \exp{\left\{-X
\frac{1+ \mathrm{e}^{-\beta\omega_D}}{1-\mathrm{e}^{-\beta\omega_D}}(1\pm \cos \theta)\right\}} \cdot \cos(Y \theta\mp X \sin \theta)
\end{align}
Defining $f_\beta \equiv
\frac{1+ \mathrm{e}^{-\beta\omega_D}}{1-\mathrm{e}^{-\beta\omega_D}}\geq 1$, the only non-trivial limit would then be $X, Y \gg 1$ for $F_{\beta,+}$. In this case, we can estimate:
\begin{align}
F_{\beta,+} &= \frac{Y}{\sin(\pi Y)} \int_{0}^{\pi} \mathrm{d} \theta \exp{\left\{-X
f_\beta(1 - \cos \theta)\right\}} \cdot \cos(Y \theta + X \sin \theta - \pi Y) \nonumber\\
&\approx Y\int_{0}^{\infty} \mathrm{d} \theta \exp{\left\{-\frac{X}{2}
f_\beta\theta^2\right\}} \cdot \left[\sin(X\theta+Y\theta) + \cos(X\theta+Y\theta) \cdot \cot(\pi Y) \right]  \nonumber\\
&= \frac{Y}{\sqrt{X f_\beta/2}} \cdot D(\frac{X+Y}{\sqrt{2X f_\beta}})   \ \ \ \ \text{D denotes Dawson's integral} \nonumber\\
&  \ \ \ \ \ + \cot(\pi Y) \cdot \sqrt{\frac{\pi}{2Xf_\beta}}Y\mathrm{e}^{-\frac{(X+Y)^2}{2Xf_\beta}} \nonumber\\
&\approx \frac{Y}{X+Y} + \frac{XY f_\beta}{(X+Y)^3} + \mathcal{O}(1/(X+Y)^3) \\
F'_{\beta,+} &= \frac{\partial F_{\beta,+}}{\partial Y}\bigg{|}_{Y\rightarrow0} \approx 1 + \frac{f_\beta}{X} + \mathcal{O}(1/X^2)
\end{align}
A similar calculation shows that $F_{\beta,-}$ and $F'_{\beta,-}$ are exponentially suppressed by $\mathrm{e}^{-\frac{(X+Y)^2}{2Xf_\beta}} $ and $\mathrm{e}^{-\frac{X}{2f_\beta}} $. The above estimations are valid as long as $Y$ is not close to any integer. For the relatively simple case $X + Y\ll 1$, it is also easy to show that $f_\beta$ starts to appear from the first order in $X,Y$.

Therefore, as long as $X, Y \gg (f_\beta-1)$ or $\ll 1/(f_\beta-1)$, our calculation of coefficients in the effective theory shall be accurate. For convenience we define $\epsilon_\beta \equiv f_\beta - 1 = f_\beta \equiv
\frac{2\mathrm{e}^{-\beta\omega_D}}{1-\mathrm{e}^{-\beta\omega_D}}$

\subsection{E. Higher Order Interaction}

We consider the virtual process amplitude in which an electron hops around a $m$-site ring as representative for higher order interactions. Taking $U_\text{eff}>0$ case as example, this process represents the superexchange among $m$ singly occupied sites, the interaction strength related to this amplitude can be written as:
\begin{align}
J^{(m)} & = \frac{t^m}{U^{m-1}_\text{eff}} F^{(m)}(X,Y) \nonumber\\
F^{(m)}(X,Y) & = Y^{m-1} \sum^{\infty}_{n_0,n_1,\dots,n_{m-1} =0} \langle 0, \dots, 0|  \hat{S}_{0,1} \frac{|n_0, n_1, 0, \dots, 0\rangle \langle n_0, n_1, 0, \dots, 0| }{(n_0+n_1)+Y} \cdots \nonumber \\
& \ \ \ \ \ \ \ \ \ \ \ \ \ \ \ \ \ \ \ \ \ \ \ \ \ \ \ \  \ \ \ \frac{|n_0,0, \dots, 0, n_{m-1}\rangle \langle n_0,0, \dots, 0, n_{m-1}| }{(n_0+n_{m-1})+Y} \hat{S}_{m-1,0}  | 0, \dots, 0 \rangle  \nonumber\\
& = Y^{m-1} \int_0^\infty \mathrm{d}t_1\cdots \mathrm{d}t_{m-1} \sum^{\infty}_{n_0,n_1,\dots,n_{m-1} =0} \exp \left[- (n_0+Y) \sum^{m-1}_{i=1} t_i -  \sum^{m-1}_{i=1} n_i t_i - \frac{mX}{2} \right] \prod_{i=1}^{m-1} \frac{(\frac{X}{2})^{n_i}}{n_i!} \nonumber\\
& = Y^{m-1} \mathrm{e}^{-\frac{mX}{2}} \int_0^\infty \mathrm{d}t_1\cdots \mathrm{d}t_{m-1} \exp \left[- Y \sum^{m-1}_{i=1} t_i + \frac{X}{2}\left(\mathrm{e}^{-\sum^{m-1}_{i=1} t_i} + \sum^{m-1}_{i=1}\mathrm{e}^{- t_i}\right) \right] \\
& \leq  Y^{m-1} \mathrm{e}^{-\frac{mX}{2}} \int_0^\infty \mathrm{d}t_1\cdots \mathrm{d}t_{m-1} \exp \left[- Y \sum^{m-1}_{i=1} t_i + r_m X\left(\sum^{m-1}_{i=1}\mathrm{e}^{- t_i}\right) \right] \nonumber\\
& = [F(r_m X,Y)]^{m-1} \ \ \ \ \  (r_m \equiv\frac{m}{2(m-1)}  )
\end{align}
With the aid of approximation in Eq.~\ref{approx}, we can prove that a sufficient condition for $J^{(2)}\gg J^{(m)}$ is (assuming $U_\text{e-e} \gg U_\text{eff} \sim t$): $U_\text{eff} \gg t$ or  $|U_\text{eff}|\gg r_m  U_\text{e-ph} \mathrm{e}^{-X}$. A nearly identical calculation for $V_b$ yields the same condition of $|U_\text{eff}|$.

The corresponding phonon-assisted hopping process amplitude can be seen as a representative of  $V^{(m)}$. It can be written as $V^{(m)} = \frac{t^m}{U^{m-1}_\text{e-ph}} F'^{(m)}(X)$ and $F'^{(m)}(X) =  X^{m-1} \partial_Y^{m-1}F^{(m)}(X,Y) |_{Y\to 0} /(m-1)! $. Equivalently, we can also compute by eliminating the $1/Y$  divergence order by order when taking $Y\to 0$ limit. This leads to: 
\begin{align}
J^{(m)}& = \frac{t^m}{U_\text{e-ph}^{m-1}}F'^{(m)}(X) \nonumber\\
F'^{(m)}(X) & = X^{m-1} \mathrm{e}^{-\frac{mX}{2}} \int_0^\infty \mathrm{d}t_1\cdots \mathrm{d}t_{m-1} \left[\exp \left(\frac{X}{2}\mathrm{e}^{-\sum^{m-1}_{i=1} t_i} \right) - 1 \right] \exp
\left[ \frac{X}{2} \sum^{m-1}_{i=1}\mathrm{e}^{- t_i} \right]  +\mathrm{e}^{-\frac{X}{2}} [F'(X/2)]^{m-1}
\end{align}
When $X\ll 1$, $F'^{(m)}(X)\to \frac{X^m}{2}+(\frac{X}{2})^{2(m-1)}$; when $X \gg 1$, dominant contribution comes from the region $t_i<\frac{1}{X}$, so $F'^{(m)}(X)\to 1$. Therefore, a sufficient condition for $J^{(2)}\gg J^{(m)}$ is $U_\text{e-ph} \gg t\cdot \text{min}(X\cdot2^{-\frac{1}{m-2}},1)$.

\subsection{F. Small clusters in $|t|\ll J, V$ model}

Dilute collection of electrons tend to form small clusters in $|t|\ll J, V$ model. The shape of the clusters is energetically optimized over all possible configurations. In Ref.~\cite{PhysRevB.42.6523} the authors only considered square lattice, here we extend the exact diagonalization calculation to triangular, honeycomb and Kagome lattices and list the per-particle energies for various candidate clusters. Other clusters with clearly higher energy (e.g. longer rings of Heisenberg chain~\cite{PhysRevB.48.6141}) or size larger than $16$ are not shown or considered. 
\begin{center}
\begin{tabular}{ c | c | c | c | c}
 & dimer & tetramer & hexamer & infinite \\
square lattice & \multirow{4}{*}{$\frac{1}{2}(V-J)$} & $V-\frac{3}{4}J$ & $\times$ & $2V-1.168J$\\
triangular lattice & ~ & $\frac{5}{4}V-\frac{3}{4}J$ &  \multirow{3}{*}{$V - 0.7171J$} &$3V-1.296J$~\cite{PhysRevLett.82.3899}\\
honeycomb lattice & ~ & $\times$ & ~ & $1.5V- 0.9195J$\cite{Reger_1989}\\
Kagome lattice&~&$\times$&~&$2V-0.9322J$
\cite{PhysRevLett.104.187203}
\end{tabular}
\end{center}
In all these lattices, monomer is optimal for $V>J$. Lowering the repulsion, dimer becomes favorable. Further reducing $V$, different lattices lead to different results. On square lattice, tetramer minimizes energy for $V/J < 1/2$ until phase separation occurs at $V/J < 0.418$. On triangular/honeycomb/Kagome lattice, hexamer minimizes energy for $V/J<0.434$ until phase separation occurs at $0.290$/$0.390$/$0.215$.

\subsection{G. Singlet pairing on triangular lattices}

\begin{figure}[t]
\begin{subfigure}
  \centering
  \includegraphics[width=0.38\textwidth]{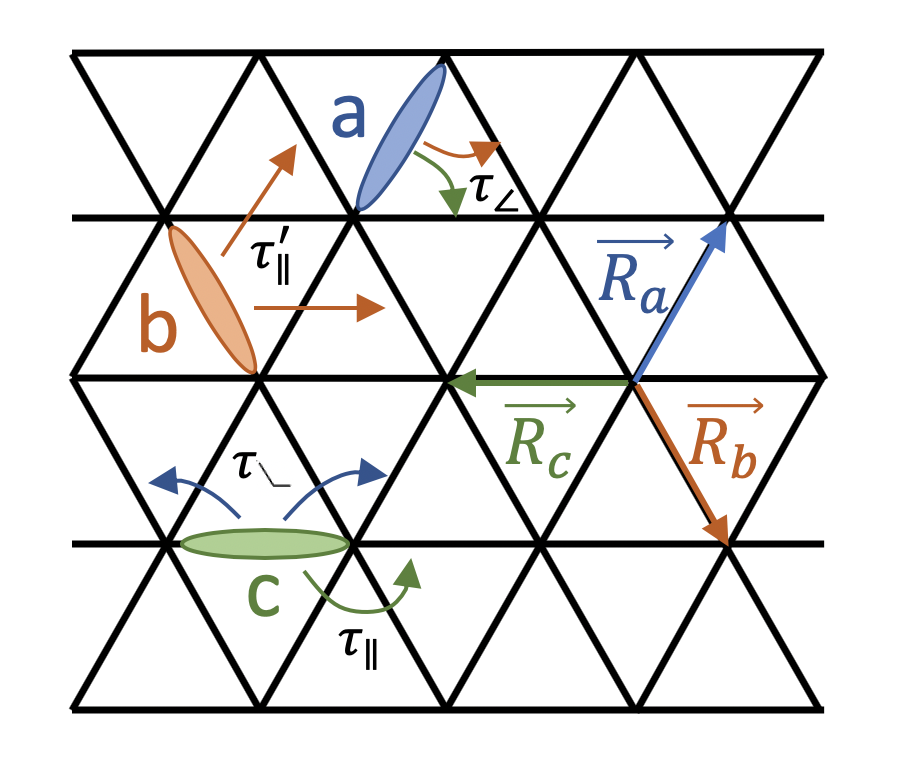}
\end{subfigure}
\hfill
\begin{subfigure}
\centering
  \includegraphics[width=0.5\textwidth]{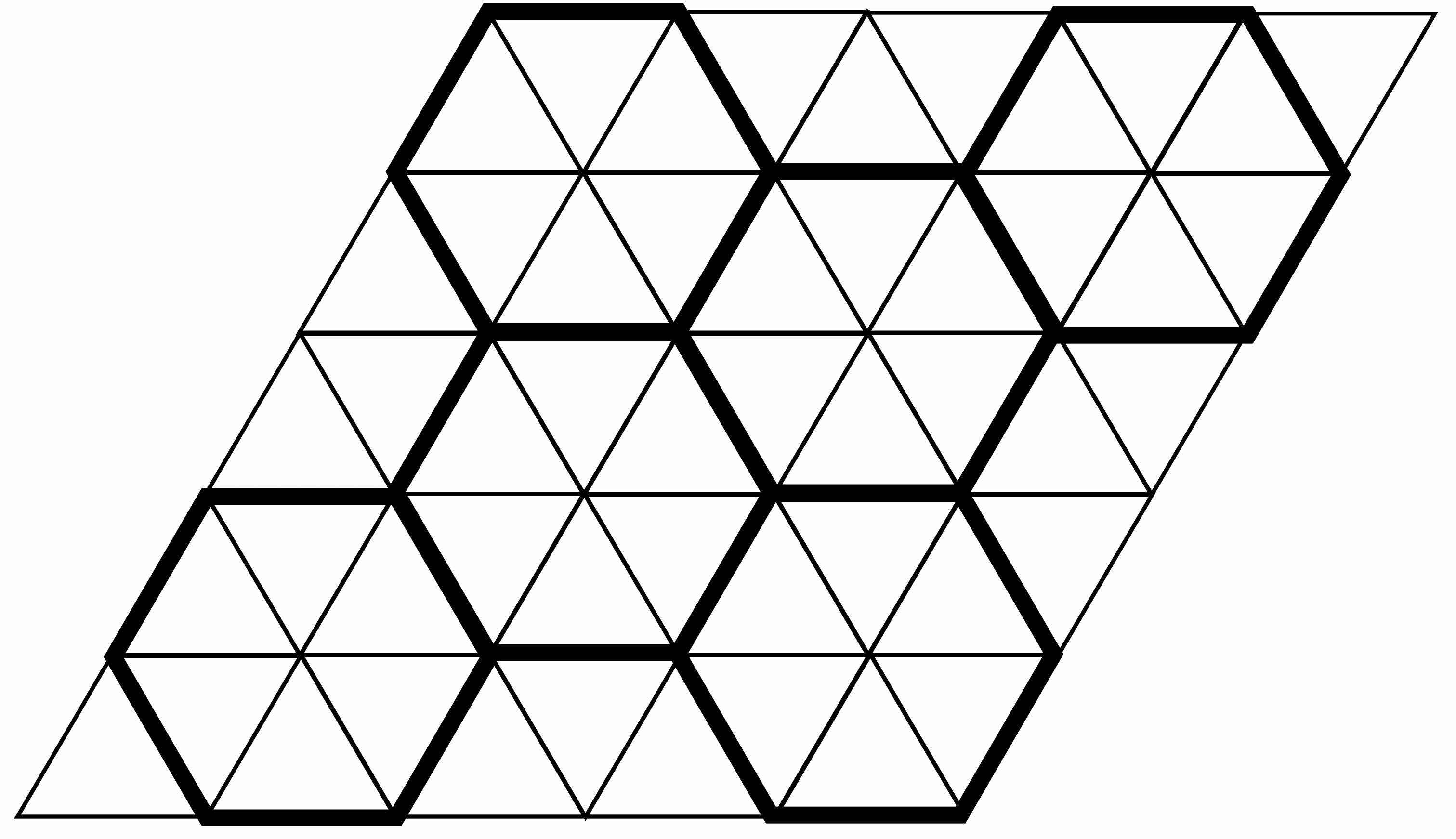}
\end{subfigure}
  \caption{(a) The various pair hopping terms on the triangular lattice.  (b) A possible PDW configuration with time reversal symmetry. }\label{triangular}
\end{figure}

For the Holstein-Hubbard model with $U_\mathrm{eff}>0$, the low-energy effective Hamiltonian is a $t-J-V$ model (leaving implicit projection onto the space of no doubly-occupied sites and hermitian conjugation of quantum hopping terms)
\begin{align}
\hat{H}_{\text{eff}} =& - t_1 \sum_{\langle i,j\rangle,\sigma} \hat{c
}_{i,\sigma}^{\dagger}  \hat{c}_{j,\sigma} + J\sum_{\langle i,j\rangle }\left[\Vec{{S}}_i\cdot\Vec{{S}}_j - \frac{\hat{n}_i \hat{n}_j}{4} \right] + V \sum_{\langle i,j\rangle } \hat{n}_{i} \hat{n}_{j}  \nonumber \\
& - \sum_{\langle i, m, j \rangle } \left[ t_2 \sum_{\sigma}  \hat{c}_{i,\sigma}^{\dagger}(1-2\hat n_m ) \hat{c}_{j,\sigma} + (\tau+2t_2) \hat{s}^{\dagger}_{mi} \hat{s}_{mj}  \right]
\end{align}
In the adiabatic limit of $\omega_{D}\to 0$ ($X, Y\to\infty$, $t_1,t_2,\tau \to 0$) with suitable range of $1>V/J>1/2$, dilute electrons form singlet dimers on nearest neighboring bonds due to the dominance of interactions. Taking back the quantum hopping terms into account, it is straightforward to obtain the effective model for a single dimer that hints the ground state pairing:
\begin{eqnarray}
H_\mathrm{dimer}=-\sum_{\langle ij\rangle ,\langle mn\rangle} \left(\tau_{ij,mn}\hat s^\dag_{ij} \hat s_{mn} +\textrm{h.c.}\right),
\end{eqnarray}
where $\tau_{ij,mn}$ is the effective pair hopping amplitude between bond $\langle ij\rangle$ and bond $\langle mn\rangle$, $\hat s^\dag_{ij}$ is the singlet creation operator on bond $\langle ij\rangle$. To the linear order of $t_1$,
$\tau$ and $t_2$ and to the second order of $t_1$, there are four types of pair hopping. On triangular lattice, as shown in Fig.~\ref{triangular}, the four types of hopping have amplitudes:
\begin{eqnarray}
&&\tau_\parallel=\tau+ t_2+\frac{t_1^2}{J-V}, \\
&&\tau'_\parallel=\frac{t_1^2}{J-V},     \\
&& \tau_{\obtuseangle{}}=\tau+2t_2+\frac{t_1^2}{J-V}\\
&&\tau_\angle{}=t_1+\tau_{\obtuseangle{}}.
\end{eqnarray}

In each unit cell, there are three possible states for dimers on a,b,c-bonds. In the momentum space, we obtain the $3\times 3$ pair hopping matrix $h_{\vec k}$:
\begin{eqnarray}
\frac{1}{2}h_{\vec k}=-\left(\begin{array}{ccc}
             \tau_\parallel \cos k_a+\tau'_\parallel(\cos k_b+\cos k_c) & \tau_{\angle{}} \cos\frac{k_c}{2}+\tau_{\obtuseangle{}}\cos\frac{k_a-k_b}{2} & \tau_{\angle{}} \cos\frac{k_b}{2}+\tau_{\obtuseangle{}}\cos\frac{k_a-k_c}{2}  \\
             \tau_{\angle{}} \cos\frac{k_c}{2}+\tau_{\obtuseangle{}}\cos\frac{k_a-k_b}{2} & \tau_\parallel \cos k_b+\tau'_\parallel(\cos k_a+\cos k_c)  & \tau_{\angle{}} \cos\frac{k_a}{2}+\tau_{\obtuseangle{}}\cos\frac{k_b-k_c}{2} \\             \tau_{\angle{}} \cos\frac{k_b}{2}+\tau_{\obtuseangle{}}\cos\frac{k_a-k_c}{2} & \tau_{\angle{}} \cos\frac{k_a}{2}+\tau_{\obtuseangle{}}\cos\frac{k_b-k_c}{2} & \tau_\parallel \cos k_c+\tau'_\parallel(\cos k_a+\cos k_b)
           \end{array}\right).
\end{eqnarray}
where $k_i\equiv \vec{k}\cdot \vec{R}_i$, and $\vec{R}_{a,b} \equiv (1,\pm \sqrt{3})/2$, $\vec{R}_c= (-1,0)$.

In general the eigenvalues are hard to acquire. Through numerical experiments, we find the minima lie along high symmetry line $k_y=0$. Then we consider:
\begin{eqnarray}
\frac{1}{2}h_{\vec k}\big{|}_{k_y=0}=-\left(\begin{array}{ccc}
             \tau_\parallel \cos \frac{k_x}{2}+\tau'_\parallel(\cos k_x+\cos \frac{k_x}{2}) & \tau_{\angle{}} \cos\frac{k_x}{2}+\tau_{\obtuseangle{}} & \tau_{\angle{}} \cos\frac{k_x}{4}+\tau_{\obtuseangle{}}\cos\frac{3k_x}{4}  \\
             \tau_{\angle{}} \cos\frac{k_x}{2}+\tau_{\obtuseangle{}} & \tau_\parallel \cos \frac{k_x}{2} +\tau'_\parallel(\cos k_x+\cos \frac{k_x}{2})  & \tau_{\angle{}} \cos\frac{k_x}{4}+\tau_{\obtuseangle{}}\cos\frac{3k_x}{4} \\ 
             \tau_{\angle{}}\cos\frac{k_x}{4}+\tau_{\obtuseangle{}}\cos\frac{3k_x}{4} & \tau_{\angle{}} \cos\frac{k_x}{4}+\tau_{\obtuseangle{}}\cos\frac{3k_x}{4} & \tau_\parallel \cos k_x+2\tau'_\parallel\cos\frac{k_x}{2}
           \end{array}\right).
\end{eqnarray}
The eigenvalues are
\begin{align}
\epsilon_0(k_x) =& -2[\tau'_\parallel \cos k_x+ (\tau_\parallel + \tau'_\parallel - \tau_{\angle{}}) \cos \frac{k_x}{2}-\tau_{\obtuseangle{}} ]\\
\epsilon_\pm(k_x) =& -[(\tau'_\parallel + \tau_\parallel)\cos k_x+ (\tau_\parallel + 3\tau'_\parallel + \tau_{\angle{}}) \cos \frac{k_x}{2}+\tau_{\obtuseangle{}}] \nonumber\\
&\pm \sqrt{\left[(\tau'_\parallel - \tau_\parallel)\cos k_x+ (\tau_\parallel -\tau'_\parallel + \tau_{\angle{}}) \cos \frac{k_x}{2}+\tau_{\obtuseangle{}}\right]^2 + 8(\tau_{\angle{}}\cos\frac{k_x}{4}+\tau_{\obtuseangle{}}\cos\frac{3k_x}{4})^2}
\end{align}

For lower band $\epsilon_-$, there is a minimum at $k_x=\frac{4\pi}{3}$ with $\epsilon_-(\frac{4\pi}{3}) = \frac{1}{2} (\tau_{\angle{}}-2 \tau_{\obtuseangle{}}) - 3/2|\tau_{\angle{}} - 2 \tau_{\obtuseangle{}}| + \tau_\parallel + 2 \tau'_\parallel$, and an extremum at $k_x=0$ with $\epsilon_-(0) =  -(\tau_{\angle{}}+ \tau_{\obtuseangle{}}) + 3|\tau_{\angle{}}+ \tau_{\obtuseangle{}}| -2 \tau_\parallel -4 \tau'_\parallel$. When $\epsilon_-(\frac{4\pi}{3})<\epsilon_-(0)$ (which requires $-t_1>2\tau +4t_2+\frac{2t_1^2}{J-V}$ and $\frac{t_1^2}{J-V}<\tau +3t_2$, and is always true in the adiabatic limit $X\to \infty$), there is a PDW with wavevector on K and -K points ($\Vec{k}=(\pm\frac{4\pi}{3},0)$). The corresponding eigenvector is $(\psi_a,\psi_b,\psi_c) = (-1,-1,1)$ and a possible PDW pattern with time reversal symmetry can be acquired, as shown in Fig.~\ref{triangular}(b), where the singlets equally condense to K and -K points.

\end{document}